\RequirePackage{rotating}
\documentclass[trackchanges,twocolumn,table]{aastex63}
\usepackage{xcolor}
\usepackage{epsfig}
\usepackage{epstopdf}
\usepackage{amsmath}
\usepackage{physics}
\usepackage{graphicx,textcomp,fancyhdr,hyperref,nicefrac,float,multirow,rotating,diagbox,array}
\newcolumntype{P}[1]{>{\centering\arraybackslash}p{#1}}
\submitjournal{ApJ}
\accepted{2021 July 15}

\shorttitle{The Sub-Neptune Wolf 503 b}
\shortauthors{Polanski et al.}

\definecolor{orange}{rgb}{0.8,0.4,0}

\newcommand\kms{\ifmmode{\rm km\thinspace s^{-1}}\else km\thinspace s$^{-1}$\fi}
\newcommand\ms{\ifmmode{\rm m\thinspace s^{-1}}\else m\thinspace s$^{-1}$\fi}
\newcommand\vsini{\ifmmode{v\sin{i_\star}}\else $v\sin{i_\star}$\fi}
\newcommand\logg{\ifmmode{\log{g}}\else $\log{g}$\fi}
\newcommand\teff{\ifmmode{T_{\rm eff}}\else $T_{\rm eff}$\fi}

\newcommand{\water}{\mbox{H$_2$O}~}
\newcommand{\mearth}{\mbox{~M$_{\Earth}$}~}
\newcommand{\rearth}{\mbox{~R$_{\Earth}$}~}
\newcommand{\densityunits}{\mbox{~$\text{ g cm}^{-3}$}}
\newcommand{\mass}{\ensuremath{6.26}}
\newcommand{\umass}{\ensuremath{^{+0.69}_{-0.70}}}
\newcommand{\radius}{\ensuremath{2.043}}
\newcommand{\uradius}{\ensuremath{\pm0.069}~}
\newcommand{\density}{\ensuremath{2.92}}
\newcommand{\udensity}{\ensuremath{^{+0.50}_{-0.44}}}

\newcommand{\waterfracshort}{\ensuremath{45}}
\newcommand{\uwaterfracshort}{\ensuremath{^{+19}_{-16}}}

\newcommand{\hhefrac}{\ensuremath{0.49}}
\newcommand{\uhhefrac}{\ensuremath{\pm0.28}}

\newcommand{\hhefracshort}{\ensuremath{0.5}}
\newcommand{\uhhefracshort}{\ensuremath{\pm0.3}}

\DeclareUnicodeCharacter{2212}{-} 
\begin{document}

\title{Wolf 503 b: Characterization of a Sub-Neptune Orbiting a Metal-Poor K Dwarf}

\correspondingauthor{Alex S.\ Polanski}
\email{aspolanski@ku.edu}

\author[0000-0001-7047-8681]{Alex S. Polanski} 
\affil{Department of Physics and Astronomy, University of Kansas, Lawrence, KS 66045, USA}

\author{Ian J.~M.\ Crossfield}
\affil{Department of Physics and Astronomy, University of Kansas, Lawrence, KS 66045, USA}
\nocollaboration{2}

\author[0000-0002-0040-6815]{Jennifer~A.~Burt}
\affiliation{Jet Propulsion Laboratory, California Institute of Technology, 4800 Oak Grove Drive, Pasadena, CA 91109, USA}

\author{Grzegorz Nowak}
\affiliation{Instituto de Astrofísica de Canarias, Vía Láctea s/n, 38205 La
Laguna, Tenerife, Spain}
\affiliation{Departamento de Astrofísica, Universidad de La Laguna, Spain}

\author[0000-0003-3204-8183]{Mercedes L\'{o}pez-Morales}
\affiliation{Center for Astrophysics ${\rm \mid}$ Harvard {\rm \&} Smithsonian, 60 Garden Street, Cambridge, MA 02138, USA}

\author[0000-0001-7254-4363]{Annelies Mortier}
\affiliation{Astrophysics Group, Cavendish Laboratory, University of Cambridge, J.J. Thomson Avenue, Cambridge CB3 0HE, UK}
\affiliation{Kavli Institute for Cosmology, University of Cambridge, Madingley Road, Cambridge CB3 0HA, UK}

\author{Ennio Poretti}
\affiliation{Fundaci\'{o}n Galileo Galilei -- INAF, Rambla Jos\'{e} Ana Fernandez P\'{e}rez 7, 38712 -- Bre\~{n}a Baja, Spain}

\nocollaboration{5}

\author[0000-0003-0012-9093]{Aida Behmard}
\affiliation{Division of Geological and Planetary Sciences, California Institute of Technology, Pasadena, CA 91125, USA}

\author{Bj\"orn Benneke}
\affiliation{Departement de Physique, and Institute for Research on Exoplanets, Universite de Montreal, Montreal, H3T J4, Canada}

\author{Sarah Blunt}
\affiliation{Department of Astronomy, California Institute of Technology, Pasadena, CA 91125, USA}

\author{Aldo S. Bonomo}
\affiliation{INAF - Osservatorio Astrofisico di Torino, Strada Osservatorio 20, I-10025 Pino Torinese (TO), Italy}

\author[0000-0003-1305-3761]{R.~Paul~Butler}
\affiliation{Earth \& Planets Laboratory, Carnegie Institution for Science, 5241 Broad Branch Road, NW, Washington, DC 20015, USA}

\author[0000-0003-1125-2564]{Ashley Chontos}
\altaffiliation{NSF Graduate Research Fellow}
\affiliation{Institute for Astronomy, University of Hawai'i, 2680 Woodlawn Drive, Honolulu, HI 96822, USA}

\author{Rosario Cosentino}
\affiliation{Fundaci\'{o}n Galileo Galilei -- INAF, Rambla Jos\'{e} Ana Fernandez P\'{e}rez 7, 38712 -- Bre\~{n}a Baja, Spain}
\affiliation{INAF - Osservatorio Astrofisico di Catania, Catania, Italy}

\author[0000-0002-5226-787X]{Jeffrey~D.~Crane}
\affiliation{The Observatories of the Carnegie Institution for Science, 813 Santa Barbara Street, Pasadena, CA 91101, USA}

\author{Xavier Dumusque}
\affiliation{Observatoire de Gen\`eve, 51 Chemin de Pegasi, 1290 Versoix, Switzerland}

\author{Benjamin J.\ Fulton}
\affil{Caltech/IPAC-NASA Exoplanet Science Institute, 770 S. Wilson Ave, Pasadena, CA 91106, USA}

\author{Adriano Ghedina}
\affiliation{Fundaci\'{o}n Galileo Galilei -- INAF, Rambla Jos\'{e} Ana Fernandez P\'{e}rez 7, 38712 -- Bre\~{n}a Baja, Spain}

\author{Varoujan Gorjian}
\affiliation{Jet Propulsion Laboratory, California Institute of Technology, 4800 Oak Grove Drive, Pasadena, CA 91109, USA}

\author[0000-0003-4976-9980]{Samuel K. Grunblatt}
\altaffiliation{Kalbfleisch Fellow}
\affiliation{American Museum of Natural History, 200 Central Park West, Manhattan, NY 10024, USA}
\affiliation{Center for Computational Astrophysics, Flatiron Institute, 162 5$^\text{th}$ Avenue, Manhattan, NY 10010, USA}

\author{Avet Harutyunyan}
\affiliation{Fundaci\'{o}n Galileo Galilei -- INAF, Rambla Jos\'{e} Ana Fernandez P\'{e}rez 7, 38712 -- Bre\~{n}a Baja, Spain}

\author[0000-0001-8638-0320]{Andrew W.\ Howard}
\affil{Department of Astronomy, California Institute of Technology, Pasadena, CA 91125, USA}

\author{Howard Isaacson}
\affil{Department of Astronomy, University of California, Berkeley, CA 94720}

\author{Molly R.\ Kosiarek}
\altaffiliation{NSF Graduate Research Fellow}
\affil{Department of Astronomy and Astrophysics, University of California, Santa Cruz, CA 95064, USA}

\author{David W.\ Latham}
\affiliation{Center for Astrophysics ${\rm \mid}$ Harvard {\rm \&} Smithsonian, 60 Garden Street, Cambridge, MA 02138, USA}

\author{Rafael Luque}
\affiliation{Instituto de Astrofísica de Canarias, Vía Láctea s/n, 38205 La Laguna, Tenerife, Spain}
\affiliation{Departamento de Astrofísica, Universidad de La Laguna, Spain}

\author{Aldo F. Martinez Fiorenzano}
\affiliation{Fundaci\'{o}n Galileo Galilei -- INAF, Rambla Jos\'{e} Ana Fernandez P\'{e}rez 7, 38712 -- Bre\~{n}a Baja, Spain}

\author{Michel Mayor}
\affiliation{Observatoire de Gen\`eve, 51 Chemin de Pegasi, 1290 Versoix, Switzerland}

\author[0000-0002-4535-6241]{Sean M. Mills}
\affiliation{Department of Astronomy, California Institute of Technology, Pasadena, CA 91125, USA}

\author{Emilio Molinari}
\affiliation{INAF - Osservatorio Astronomico di Cagliari, Selargius, Italy}

\author[0000-0002-4019-3631]{Evangelos Nagel}
\affiliation{Thüringer Landessternwarte Tautenburg, Sternwarte 5, 07778 Tautenburg, Germany}
\affiliation{Hamburger Sternwarte, Gojenbergsweg 112, 21029 Hamburg, Germany}

\author[0000-0003-0987-1593]{Enric Pall\'{e}}
\affiliation{Instituto de Astrofísica de Canarias, Vía Láctea s/n, 38205 La
Laguna, Tenerife, Spain}
\affiliation{Departamento de Astrofísica, Universidad de La Laguna, Spain}

\author[0000-0003-0967-2893]{Erik A.\ Petigura}
\affil{Department of Physics and Astronomy, University of California, Los Angeles, CA 90095, USA}

\author{Stephen~A.~Shectman}
\affiliation{The Observatories of the Carnegie Institution for Science, 813 Santa Barbara Street, Pasadena, CA 91101, USA}

\author[0000-0002-7504-365X]{Alessandro Sozzetti}
\affiliation{INAF - Osservatorio Astrofisico di Torino, Strada Osservatorio 20, I-10025 Pino Torinese (TO), Italy}

\author{Johanna~K.~Teske}
\affiliation{Earth \& Planets Laboratory, Carnegie Institution for Science, 5241 Broad Branch Road, NW, Washington, DC 20015, USA}

\author[0000-0002-6937-9034]{Sharon Xuesong Wang}
\affiliation{Department of Astronomy, Tsinghua University, Beijing 100084, People's Republic of China}

\author[0000-0002-3725-3058]{Lauren M. Weiss}
\affiliation{Institute for Astronomy, University of Hawai'i, 2680 Woodlawn Drive, Honolulu, HI 96822, USA}

\nocollaboration{31}

\begin{abstract}

Using radial velocity measurements from four instruments, we report the mass and density of a \radius \uradius \rearth sub-Neptune orbiting the quiet K-dwarf Wolf 503 (HIP 67285). In addition, we present improved orbital and transit parameters by analyzing previously unused short-cadence \textit{K2} campaign 17 photometry and conduct a joint radial velocity-transit fit to constrain the eccentricity at $0.41\pm0.05$. The addition of a transit observation by \textit{Spitzer} also allows us to refine the orbital ephemeris in anticipation of further follow-up. Our mass determination, \mass \umass \mearth, in combination with the updated radius measurements, gives Wolf 503 b a bulk density of $\rho = \density \udensity$ g$~\text{cm}^{-3}$. Using interior composition models, we find this density is consistent with an Earth-like core with either a substantial \water mass fraction (\waterfracshort \uwaterfracshort\%) or a modest H/He envelope (\hhefracshort \uhhefracshort\%). The low H/He mass fraction, along with the old age of Wolf 503 (11$\pm$2 Gyrs), makes this sub-Neptune an opportune subject for testing theories of XUV-driven mass loss while the brightness of its host ($J=8.3$ mag) makes it an attractive target for transmission spectroscopy.

\end{abstract}


\section{\textbf{Introduction}} \label{sec:intro}

One of the most notable discoveries in the exoplanet field is the ubiquity of not one, but two new classes of planet frequently found orbiting late-type stars with periods less than 100 days; the super Earths and sub-Neptunes. This planet sub-population, first discovered over a decade ago through Doppler surveys of the southern sky \citep{Mayor2008,Lovis2009}, has expanded dramatically with the \textit{Kepler/K2} and \textit{TESS} missions \citep{Fulton2017}. With over 2,000 confirmed planets, these missions have presented us with a diversity of worlds that we previously had not anticipated. As with any discovery, these planets have forced us to rethink and reformulate not only our theories of planet formation, but also the evolution of exoplanets and their atmospheres as well as how they are affected by their host stars.

\begin{figure}[h!]
\centering
\vspace{0.5cm}

\includegraphics[width=0.45\textwidth]{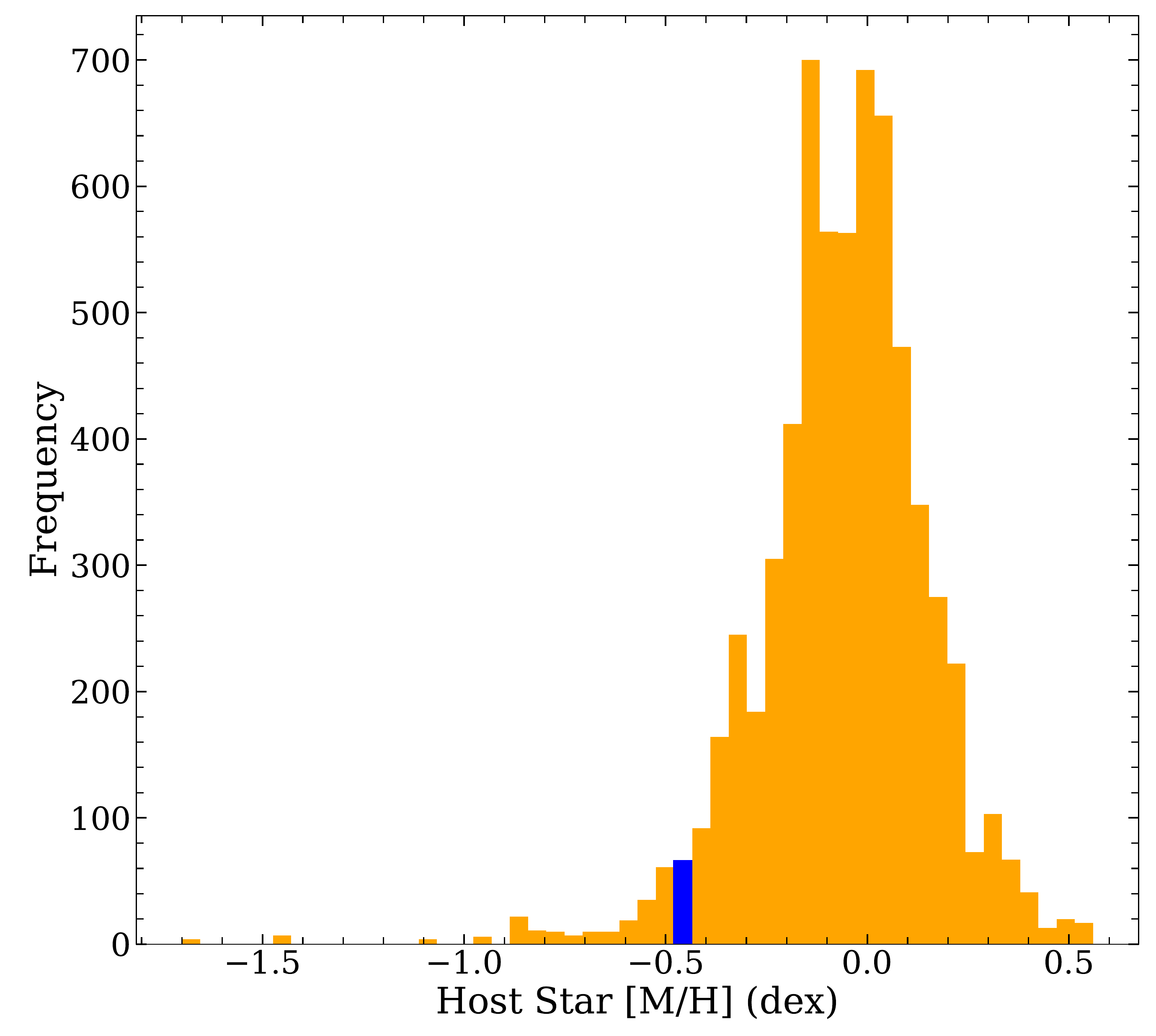}
\caption{Histogram of metallicities for all host stars with a sub-Neptune class of planet (1.6-3.2 $R_{\Earth}$). Wolf 503 has a metallicity of $\text{[M/H]}=-0.47$, finding itself in the blue shaded bin making it one of the metal poorer stars to host a sub-Neptune. Data accessed from the NASA Exoplanet Archive (Jan. 24th 2021).}
\label{fig:rad_met}
\end{figure}

By combining both radius and mass measurements with models of planetary interiors it appears that these short period sub-Neptunes (1.6-3.2 $R_{\Earth}$) potentially range from volatile-rich worlds with hydrogen/helium envelopes constituting nearly a third of their mass to rocky cores stripped of their atmosphere by their host star \citep{Rogers2010,OwenWu2013, Lopez2014}. Interior models even suggest some sub-Neptunes could host hydrospheres of super-critical water blanketed by steam-dominated envelopes \citep{Zeng2014,Thomas2016,Mousis2020}. With core-accretion as the prevailing theory of planet formation, the frequency of these planets was previously thought to be, at best, rare. The low density, high temperature environment of the proto-planetary disk within the snowline makes building planets larger than Earth through core-accretion inefficient, let alone planets with substantial gaseous envelopes. However, \textit{in-situ} formation of sub-Neptunes can still be possible as large dust grains drift from the outer disk inward and accumulate at the inner-most regions of the disk in a pebble-accretion scenario. In contrast, formation beyond the snowline and subsequent migration inward can also lead to sub-Neptunes \citep{Bodenheimer2014, Venturini2017}.

The added complexity required in our formation theories gives rise to other intriguing questions: How has planetary migration affected the exoplanet populations we observe today? Are the properties of host stars reflected in the planets that orbit them? 

In the post-\textit{Kepler} era, efforts have now shifted from discovery to characterization and the answers to those questions seem to be on the horizon. Radial velocity (RV) surveys to measure the masses of previously discovered planets have been essential in placing these planets in context. Precise masses are especially important in modeling the potential atmospheric conditions on these planets. The degeneracy between the mean molecular weight of the atmosphere (a potential indicator of metal content) and the surface gravity can have a serious impact on a planet's potential for follow-up observations since these parameters can have similar effects on transmission spectra. The best way to break the degeneracy is to have mass uncertainties $\leq$ 20\% \citep{Batalha2019}. With the James Webb Space Telescope (JWST) set to be the premier facility for studying exoplanet atmospheres, it is essential to not only measure precise masses but also constrain the transit-timing uncertainties beyond the \textit{K2} values; \textit{Spitzer} has been instrumental in assuring valuable telescope time will not be wasted on missed transit events. 

In this paper we characterize the sub-Neptune Wolf 503 b \citep{Peterson2018}. This planet orbits a bright ($J=8.3$ mag) K-dwarf star making it an intriguing candidate for future atmospheric follow-up. This work is outlined as follows: We begin by describing the properties of the host star as well as re-deriving key stellar parameters with new \textit{Gaia} Early Data Release 3 (EDR3) parallax values in Section \ref{sec:hostparams}. In Section \ref{sec:photometry} we present the analysis of \textit{K2} photometry using a photo-eccentric model which implies and eccentric orbit for Wolf 503 b. Section \ref{sec:spitzer} describes the observations made with the \textit{Spitzer} space telescope and uses the results to further constrain the orbital ephemeris. In Section \ref{sec:rv} we use radial velocity data to further confirm the planet's eccentricity and conduct a joint RV-transit fit to better constrain the orbital parameters. Section \ref{sec:discussion} discusses possible interior composition Wolf 503 b along with the potential for atmosphere characterization with JWST.

\section{\bf{Target System Parameters}}\label{sec:hostparams}

Wolf 503 (EPIC 212779563, HIP 67285) is a bright ($J=8.3$ mag) K3.5V main sequence dwarf. At 44.630 $\pm$0.033 pc, this nearby star is currently known to host one planet, Wolf 503 b, that was discovered in \textit{K2} campaign 17 photometry in  2018 \citep{Peterson2018}. Wolf 503 b was found to be a \radius \rearth planet that completes one orbit roughly every 6 days. At 0.06 AU from its star, Wolf 503 b has an equilibrium temperature $\sim$ 800 K; an intermediate temperature compared to other sub-Neptunes discovered.

\begin{deluxetable}{lcc}[h]
\tabletypesize{\footnotesize}
\setlength{\tabcolsep}{0.05cm}
\tablecaption{Stellar Parameters \label{tab:stellarparams}}
\tablehead{
  \colhead{Parameter (units)} & 
  \colhead{Value} & 
  \colhead{Provenance}
}
\startdata
EPIC ID & 212779563 & \\
$\alpha$ R.A. (hh:mm:ss) J2000 & 13:47:23.4439 & \\
$\delta$ Dec. (dd:mm:ss) J2000 & -06:08:12.731 & \\
\hline
\sidehead{\textbf{Magnitudes}}
$NUV$ (mag) & $18.521\pm0.061$ & GALEX \\
$B$ (mag) & $11.30\pm0.01$ & \cite{Mermilliod1987} \\
$V$ (mag) & $10.28\pm0.01$ & \cite{Mermilliod1987} \\
$G$ (mag) & $9.8982\pm0.0003$ & \textit{Gaia} DR2 \\
$J$ (mag) & $8.324\pm0.019$ & 2MASS \\
$H$ (mag) & $7.774\pm0.051$ & 2MASS \\
$K$ (mag) & $7.617\pm0.023$ & 2MASS \\
\hline
\sidehead{\textbf{Properties}}
$\mu_{\alpha}$ (mas yr$^{-1}$) & $-342.862\pm0.020$ & \textit{Gaia} EDR3\\
$\mu_{\delta}$ (mas yr$^{-1}$) &  $-573.112\pm0.014$ & \textit{Gaia} EDR3\\
Barycentric rv \newline (km s$^{-1}$) & $-46.826\pm0.015$ & \textit{Gaia} DR2\\
Age (Gyr) & $11\pm2$ & \cite{Peterson2018} \\
Spectral Type & $K3.5\pm0.5$V & \cite{Peterson2018}\\
$\text{[Fe/H]}$   & $-0.47\pm0.08$ & \cite{Peterson2018}\\
$\log(g)$ (K) & $4.62^{+0.02}_{-0.01}$ & \cite{Peterson2018}\\
$\teff$ (K) & $4716 \pm 60$ & \cite{Peterson2018}\\
$M_*$ ($M_{\odot}$) & $0.688^{+0.023}_{-0.016}$ & \cite{Peterson2018}\\
$R_*$ ($R_{\odot}$) & $0.689^{+0.021}_{-0.020}$ & This Work\\
$\rho_*$ (\densityunits) & $2.17 \pm 0.12$ & This Work\\
$L_*$ ($L_{\odot}$) & $0.211^{+0.007}_{-0.007}$ & This Work\\
Distance (pc) & $44.630\pm0.033$ & This Work\\
\hline
\enddata
\end{deluxetable}

In order to increase the accuracy of derived parameters such as mass and radius, we re-derived key stellar parameters for Wolf 503 using the \textit{Gaia} mission's new parallax measurements the uncertainty on which has been reduced by a factor of three \citep[EDR3][]{Gaia2020}. With values for spectroscopic parameters $\teff$, $\text{[Fe/H]}$, and $\log(g)$ from \citet{Peterson2018} and the photometric magnitude in the K-band, we use \texttt{Isoclassify} \citep{Huber2017, Berger2020} to obtain the distance, $R_*$, and $L_*$ of Wolf 503. \texttt{Isoclassify} determines stellar parameters using a sample of 2200 \textit{Kepler} stars in combination with \textit{Gaia} data with uncertainties on those parameters based on MIST data. We use the direct method described in \cite{Huber2017} to determine these parameters which are listed in Table \ref{tab:stellarparams}. The values found for distance, $R_*$, and $L_*$ agree with values previously found by \citet{Peterson2018} but the uncertainties see slight reductions ($\leq1$\% in the case of radius and luminosity). 
\subsection{A Metal-Poor Host}

The age and metallicity of its host star sets Wolf 503 b apart from the majority of sub-Neptunes. Wolf 503's age is estimated to be between 9-13 Gyrs and the star has a metallicity of $\text{[Fe/H]}=-0.47\pm0.08$ \citep{Peterson2018} making it one of the more metal poorer stars to host a sub-Neptune (Figure \ref{fig:rad_met}). It has been well established that Jupiter-class planets are frequently found orbiting stars of increasing metallicity with a correlation between close-in giant planet occurrence rate and host star metal enrichment \citep{Gonzalez1997,Santos2004,Thorngren2016}. This correlation can be understood in the context of core accretion; massive planets need more solid material in order to trigger a runaway accretion of gas. Although this trend weakens with decreasing planetary size , warm sub-Neptune occurrence is still correlated with host star metallicity \citep{Petigura2018}. However, the formation of sub-Neptune planets require specific disk conditions that balance the build up of a massive core while also preventing the runaway accretion that results in a gas giant. \cite{Venturini2017} found that formation scenarios with low solid accretion rates (~$10^{-6}M_{\Earth}yr^{-1}$) resulted in the highest sub-Neptune occurrence rate. This accretion rate is compatible with disks of low metallicites but is also possible in low mass disks as well.   

\section{\textbf{\textit{K2} Short Cadence Photometry}} \label{sec:photometry}

While the detection of eccentric orbits is usually done with radial velocity observations, through the photo-eccentric effect \citep{Dawson2012}, one can obtain broad constraints on a planet's eccentricity from its lightcurve if an independent measurement of the stellar density can be made. In this section we extract previously unused \textit{K2} short cadence photometry and, with a stellar density obtained in Section \ref{sec:hostparams}, use a photo-eccentric transit model to determine if Wolf 503 b is on a circular or eccentric orbit. 


\subsection{Lightcurve Extraction}

Wolf 503 was observed by \textit{Kepler} from 2018 March to 2018 May. We extract photometry from \textit{K2}'s target pixel file (TPF) using the \texttt{Lightkurve} package \citep{Lightkurve2018}. TPF's are the main data product of the \textit{Kepler/K2} and \textit{TESS} missions consisting of stacks of ``postage stamp'' frames centered on the target star. Each frame represents one timestamp (or cadence) in which data was taken. For \textit{Kepler/K2} short cadence, the sampling rate is about a minute between exposures whereas long cadence only samples every 30 minutes.

\begin{figure*}[t!]
\centering
\hspace*{-1.5cm}
\includegraphics[height=8.0in,width=7.0in,keepaspectratio]{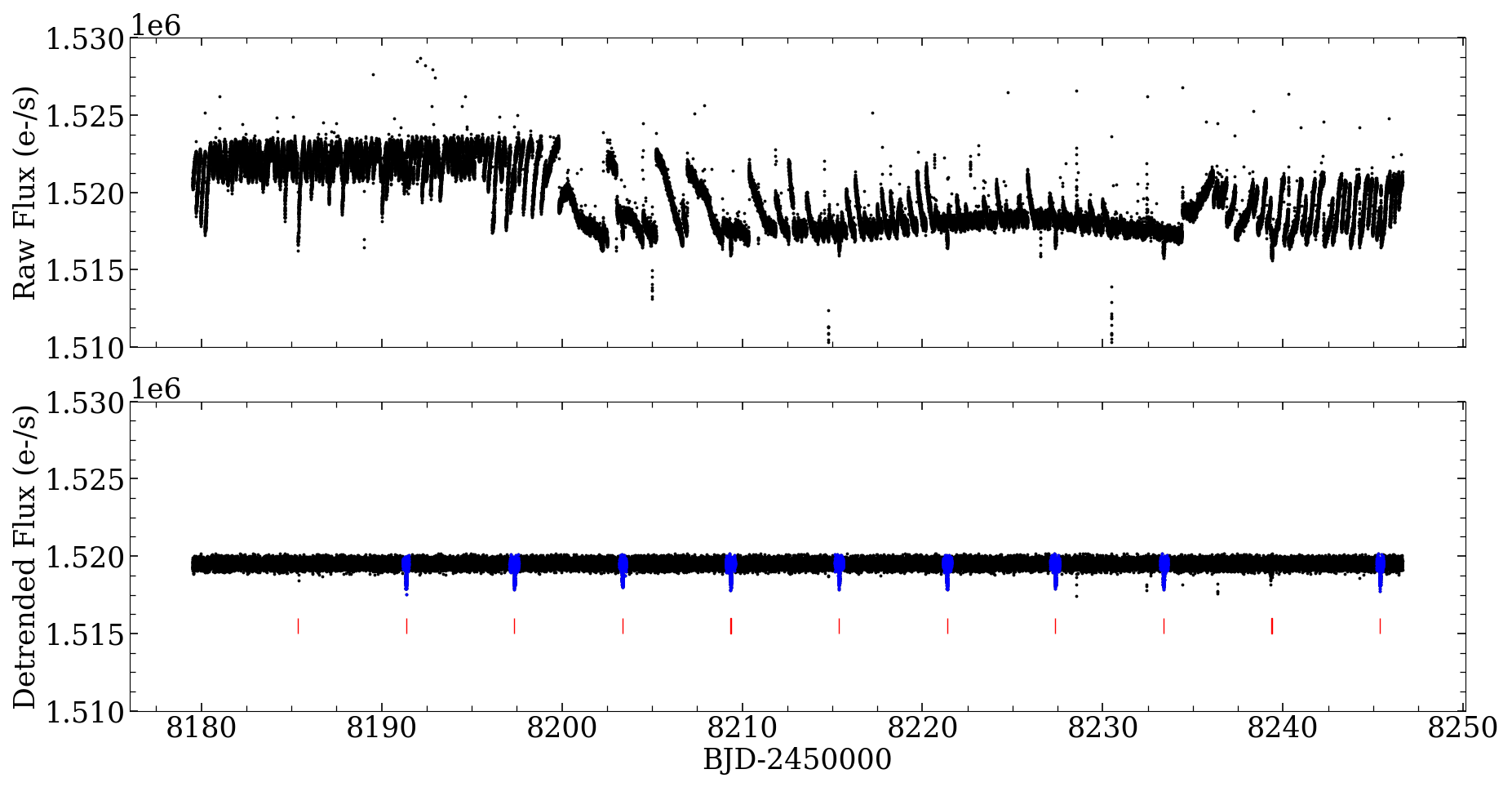} 
\caption{The top panel shows the raw lightcurve extracted from the short cadence target pixel file using aperture photometry while the bottom panel shows the detrended result after employing the Self Flat-Fielding (SFF) technique to account for \textit{Kepler's} motion during the K2 campaigns. Red tick marks indicate the transits of Wolf 503 b and the blue overlay shows which points were used in the fitting process.  The 1st and 10th transits were omitted as they coincided with thruster burns. 
}

\label{fig:extraction}

\end{figure*}

After the failure of two of \textit{Kepler's} reaction wheels, the solution that allowed \textit{K2} to be possible resulted in the target stars drifting across the detector over the length of the campaign. This drift causes changes in flux levels and needs to be corrected for. \texttt{Lightkurve} implements the Self Flat-Fielding (SFF) technique introduced by \cite{Vanderburg2014} to account for the motion of the \textit{Kepler} spacecraft. Aperture photometry was performed on the TPF using the a circular pixel mask of radius 5 pixels centered on the star. We experimented with various aperture sizes ranging from 4 to 6 pixels. The 5 pixel radius produced the lowest out-of-transit spread in the data after SFF was applied. The result of the self flat-fielding technique is show in Figure \ref{fig:extraction} with red tick marks indicating clear transit events with the exception of the first and tenth transits which suffered from thruster burns. There remains occasional decreases in flux between transits that are not periodic and are likely due to extreme differences in pixel sensitivity across the detector. Since these points are not explicitly used in fitting process they have no impact on the parameters derived in the following section.

\subsection{Photo-eccentric Model}

We fit the short-cadence data using the \texttt{exoplanet} package \citep{Foreman-Mackey2021} which uses a Hamiltonian Monte Carlo (HMC) routine to explore the posterior probability distribution. We minimize a negative log-likelihood function using the period of the orbit ($P$), time of inferior conjunction ($T_{conj}$), impact parameter ($b$), scaled planet radius ($R_p/R_*$), and stellar density assuming a circular orbit ($\rho_{*,circ}$) as free parameters and use a quadratic limb darkening law with the parameters held at $u_0=0.5916$ and $u_1=0.1322$ obtained from \cite{Claret2011}. Using the values obtained from the minimization algorithm, we initialized the HMC sampler with four parallel chains running 8,000 tuning steps and 6,000 sampling steps. Loose Gaussian priors were placed on $P$ and $T_{conj}$ and instead of sampling directly in $\rho_{*, circ}$, we reparameterize according to \citet{Sandford2017} using $\log_{10}{(\rho_{*, circ})}$ with a uniform prior.

With a median $R_p/R_*$ value of $2.534 \pm 0.020\%$, our work agrees reasonably well with the previous analysis of this system. However, we obtain $\rho_* = 16 \pm 1 \densityunits$; higher than what is expected using values found in Section \ref{sec:hostparams}. This large stellar density can be explained in terms of the photo-eccentric effect and indicates that Wolf 503b is not on a circular orbit but an eccentric one. 

While a planet on a circular orbit has a constant velocity, a planet in an eccentric orbit has a maximum velocity at periapsis and a minimum at apoapsis. This creates a dependence of the transit length on the argument of periapse ($\omega$) of the orbit. An observer viewing the transit at periapsis ($\omega = 90\degr$) would record a shorter transit than an observer at apoapsis. From \cite{winn2010} the length of the transit ($T$) can be given as,

\begin{equation} \label{eq:eq1}
    T = \qty(\frac{R_* P}{\pi a}\sqrt{(1-b^2)}) \frac{1}{g(e,\omega)}
\end{equation}

\noindent where $g$ is expressed as,

\begin{equation} \label{eq:eq2}
    g(e,\omega) = \frac{1 + e\sin{(\omega)}}{\sqrt{(1-e^2)}}
\end{equation}

\par By employing Kepler's 3rd Law, we can substitute $a/R_*$ in favor of the stellar density and obtain a key equation from \citet{Kipping2010},

\begin{equation} \label{eq:eq3}
    \rho_{*,circ} = \rho_* g(e,\omega)^3
\end{equation}

A planet on a circular orbit will give $\rho_{circ} = \rho_*$. However, a planet on an eccentric orbit, transiting near periapsis, would give a larger $\rho_{circ}$ compared to an independent measurement of the star's density.

Using the methodology of \citet{Dawson2012}, we sample values of $e$ and $\omega$ from uniform distributions on the interval $[0,1]$ and $[-\pi,\pi]$, respectively and compute the $g$ parameter using Equation \ref{eq:eq2}. These values of $g$ are then used together with the $\rho_{*,circ}$ to calculate what true density is implied from Equation \ref{eq:eq3} and is then compared to the value found with \texttt{isoclassify} using a likelihood function. The likelihoods are then used to re-weight the samples of $e$ and $\omega$ in order to obtain a an estimate of the eccentricity yielding a $1\sigma$ range of $0.59$ to $0.82$.

\begin{figure*}
\centering
\vspace{0.5cm}

\includegraphics[width=0.9\textwidth]{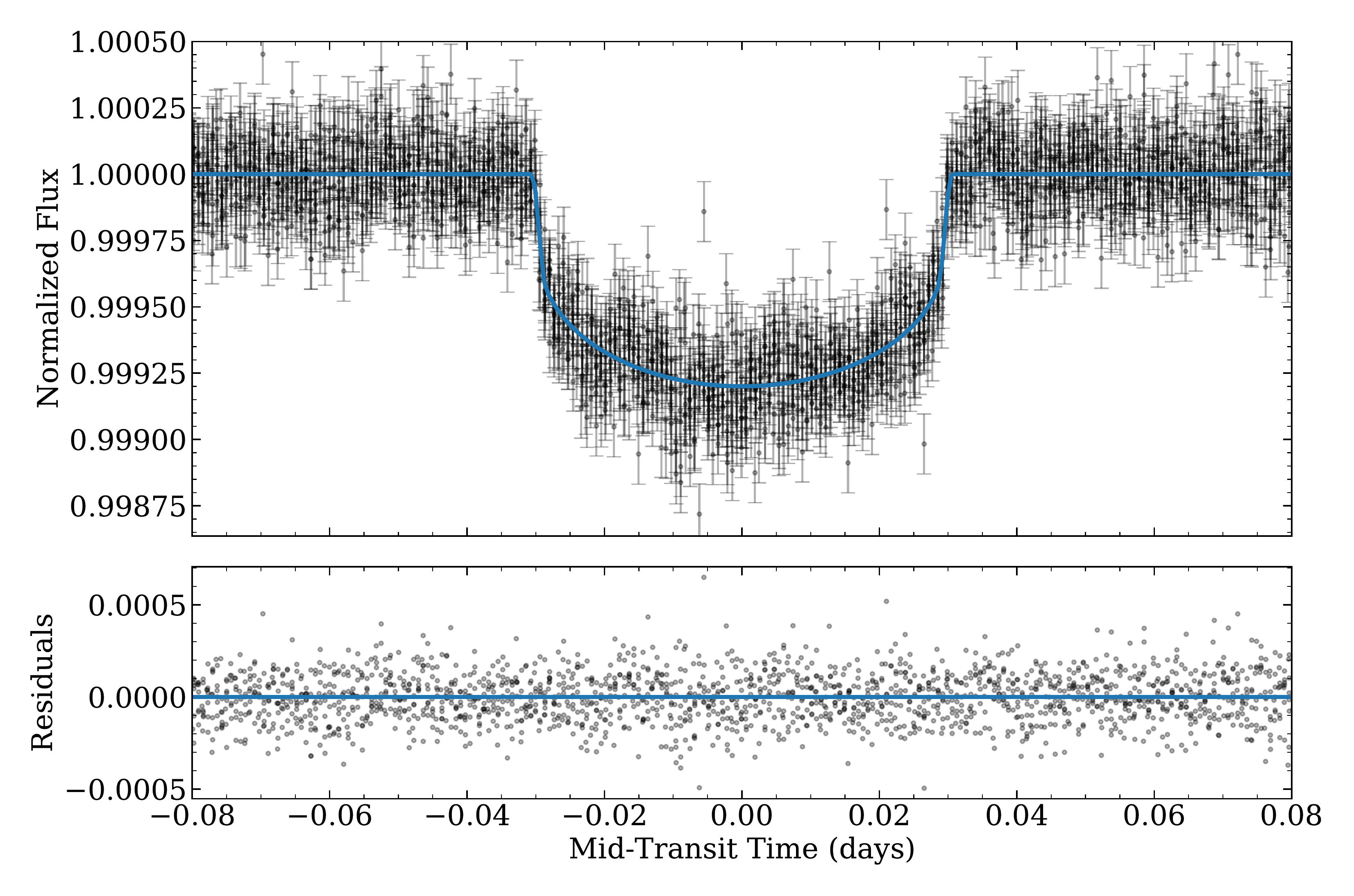}
\caption{Transit fit to \textit{K2} short-cadence photometry of Wolf 503 b. Individual data points are given with their 1-sigma uncertainties while the fit from median posterior values is shown as the blue line. }
\label{fig:k2_sc}
\end{figure*}

\section{\textbf{\textit{Spitzer} Photometry}}\label{sec:spitzer}

After the discovery of Wolf~503 b, we were awarded Director's Discretionary Time \citep{crossfield:2019ddt} to observe the planet's transit with {\em Spitzer}.  On 2019/11/10 we observed one transit using the 4.5\,\micron\ channel \citep[IRAC2,][]{fazio:2005} with 2.0\,s integrations  taken in subarray mode; the transit observation encompassed 208 frames and spanned 7\,hr 27\,min. In addition, we acquired a short observation before and after the transit to check for bad pixels. Our observations were scheduled following standard best practices for precise \textit{Spitzer} photometry, including using Peak-Up mode to place the star as closely as possible to the well-characterized ``sweet spot'' of the IRAC2 detector.

\subsection{POET Reduction Pipeline}

To extract photometry from the \textit{Spitzer} observations, we use the Photometry for Orbits Eclipses and Transits (\texttt{POET} \footnote{\url{https://github.com/kevin218/POET}}) package \citep{Cubillos2013,May2007}. In summary, \texttt{POET} creates a bad pixel mask and discards bad pixels based on the \textit{Spitzer} Basic Calibrated Data (BCD). Outlier pixels are also discarded using sigma-rejection. Then, the center of the point spread function (PSF) is determined. \textit{POET} provides multiple routines to determine the PSF center and since we see no evidence of any source near Wolf 503 (\cite{Peterson2018}) we opt for a simple 2-D Gaussian fitting technique. After the center of the PSF is found, interpolated aperture photometry is used to extract the lightcurve. The resulting data is then fit with a model that accounts for both the lightcurve itself in addition to a ramp-like trend attributed to ``charge trapping'' (discussed in \ref{sec:spitzer_systematics}) and the sub-pixel sensitivity of the detector. The posterior distribution is sampled using an MCMC algorithm with chains initialized at the best fit values.

\subsection{Interpolated Aperture Photometry}

The quality of the fit is dependent not only on the model and the aperture size, but also the method of interpolation and the bin-size used. To find the best result, we tested various aperture sizes (ranging from 2 - 6 pixels in increments of 1 pixel) with both nearest neighbor (NNI) and bilinear (BLI) interpolation using different bin sizes (0.1, 0.03, 0.01, and 0.003 square). For each case, the standard deviation of the normalized residuals (SDNR) was calculated and compared. The method resulting in the lowest SDNR was an aperture size of 5 pixels with interpolated photometry performed with bilinear interpolation and a bin size of 0.03 x 0.03 pixels.

\subsection{\textit{Spitzer} Systematics}\label{sec:spitzer_systematics}

At 4.5 $\mu$m, the primary systematic effect is the sub-pixel sensitivity variation causing the measured flux to be dependent on the target's position on the array \citep[][]{Stevenson2012,Charbonneau2005,Cubillos2013}. To mitigate this variability, we utilize the BiLinearly-Interpolated Subpixel Sensitivity (BLISS) mapping described in \cite{Stevenson2012} which has been shown to be a more effective method of mapping the sub-pixel sensitivity of the detector as compared to polynomial fits or the weighted sensitivity function of \cite{Ballard2010}.

There is also a temporal systematic that induces a ramp-like trend in the extracted lightcurve (see Figure \ref{fig:spitzer}a). The cause of this is thought to be charge trapping \citep{Agol2010} and is an issue especially for brighter targets. During read-out of the detector, not all electrons are drained from the pixel leaving to be ``trapped'' in the pixel. As the observation progresses, the electrons build up increasing the effective gain of the detector. This manifests itself as a ramp-like trend in the lightcurve. To account for this, our fitted model uses a linear trend of the form:

\begin{equation}
    R(t) = 1 +r_1(t-t_{0})
\end{equation}

Where $r_1$ is the slope of the linear model and a free parameter which is fit for. $t_0$ is a constant term approximated as the mid-point phase of the transit and, in our case, set to be 1.0.

\subsection{Lightcurve Fitting}\label{sec:spitzer_lightcurve}

The model lightcurve is generated using the \texttt{batman} package. Since we observed only one transit event with \textit{Spitzer} (as opposed to the eight analyzed from \textit{K2}) the uncertainties on transit parameters will necessarily be larger. The primary advantage of observing Wolf 503 b with \textit{Spitzer} is making accurate predictions of future transits (Section \ref{sec:ephemeris}).

\begin{figure}[htbp]
\centering
\includegraphics[width=0.5\textwidth]{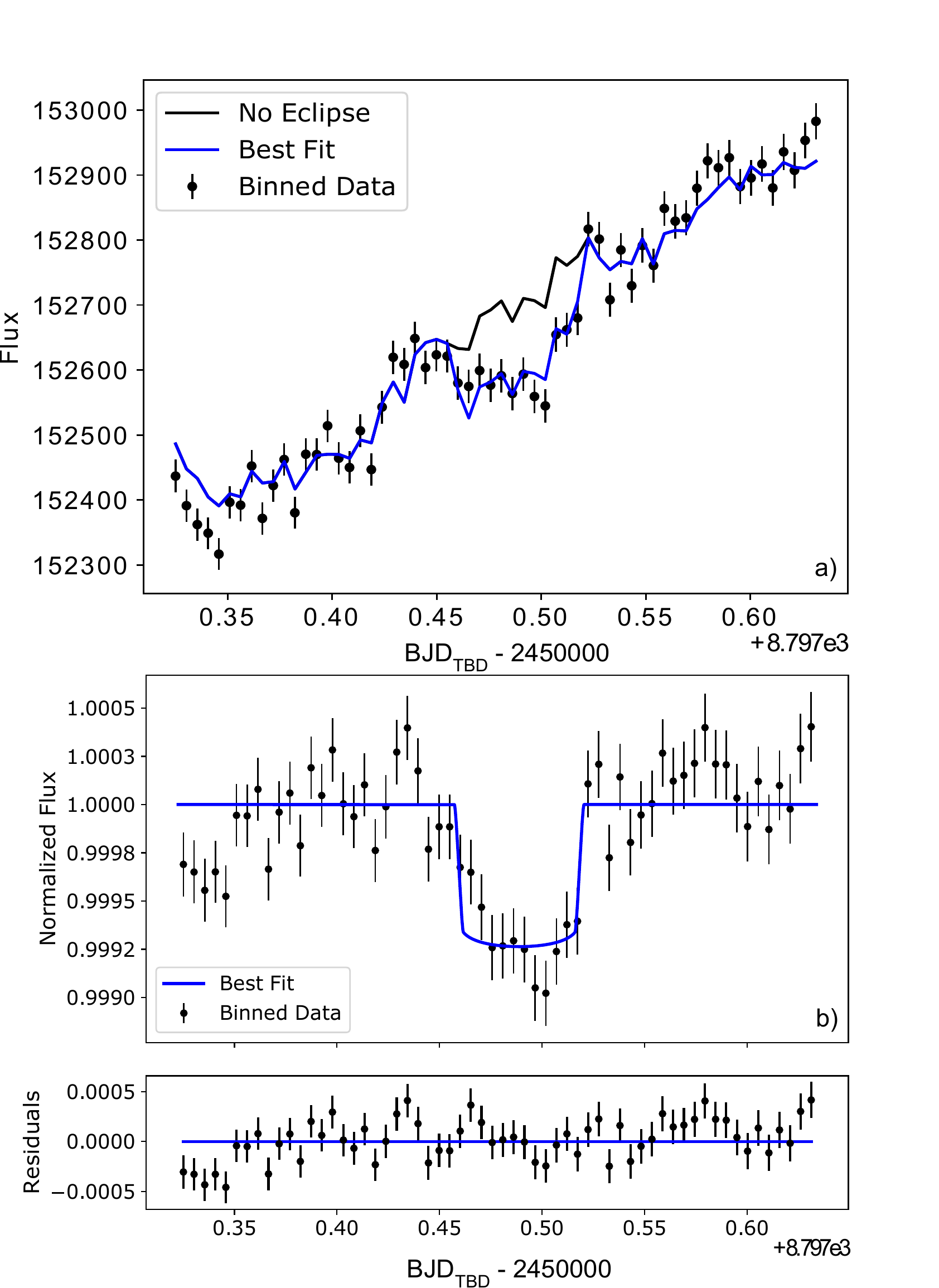}
\caption{a) Lightcurve from \textit{Spitzer} showing the ramp-like trend in flux over the course of the transit. b) \textit{Spitzer} lightcurve showing the best fit transit curve. }
\label{fig:spitzer}
\end{figure}

After performing a least squares fit, an MCMC routine is initialized on the best-fit values and allowed to go though 100,000 iterations taking the first 3,000 as burn in. The parameters involved the analysis are $R_P/R_*$, $T_{conj}$, $a/R_*$, $cos(i)$, and the parameters in Eq. \ref{eq:1}. A Gaussian prior was placed on $a/R_*$ informed by the value found from our radial velocity analysis in Section \ref{sec:rv_analysis}. We use a quadratic limb darkening law with the parameters held at $u_0=0.0973$ and $u_1=0.1276$ obtained from \cite{Claret2011}. To increase computational efficiency, the eccentricity and argument of periastron are held at values found in \ref{sec:rv_analysis}. The median posterior value for the transit depth is given in table \ref{tab:rvparams} and the lightcurve fit to the photometry is shown in Figure \ref{fig:spitzer}b.

\subsection{Ephemeris Improvements}\label{sec:ephemeris}

With \textit{K2} photometry alone, the uncertainties on a planet's period ($P$) and the time of inferior conjunction ($T_{conj}$) degrade our ability to predict transits in the future. The uncertainty in mid-transit time ($T_n$) scales linearly with the number of orbits ($n$) since the initial observation \citep{beichman2016}:

\begin{equation}
    \sigma_{T_n} = \sqrt{\sigma_{T_{conj}}^2 + (n\sigma_{P})^2} \label{eq:1}
\end{equation}

By the time JWST is operational (2022), the 3-sigma uncertainty in the transit time, calculated from the long-cadence period alone, would be nearly 2 hours. For a planet whose total transit time lasts little over an hour, there is a likely chance we would only observe a partial transit, or in the worst case, miss the transit entirely. Both of these situations are unacceptable uses of valuable telescope time making precise knowledge of when a transit will occur crucial for future follow up studies.

In order to tighten our constraint on the mid-transit time we use the conjunction time obtained with both \textit{K2} and \textit{Spitzer} and use a weighted least squares routine to obtain a more precise value of the period. A weighted least squares method is used instead of a joint fit to both the \textit{K2} and \textit{Spitzer} photometry as the systematics in the \textit{Spitzer} data tend to be so strong that it is difficult to model the systematics independently of the transit itself. We obtain a new period of $6.001274\pm2.1e-05$ days. With the period obtained from the short cadence \textit{K2} photometry analyzed in this work, the precision improves to 34 minutes and with the addition of the \textit{Spitzer} transit we ultimately come to a mid-transit time precision of just 21 minutes; a 5-fold improvement from the \textit{K2} long cadence prediction.    

\section{\textbf{Radial Velocity Analysis}}\label{sec:rv}

\begin{figure*}[!t]
\centering
\includegraphics[height=8.0in,width=6.0in,keepaspectratio]{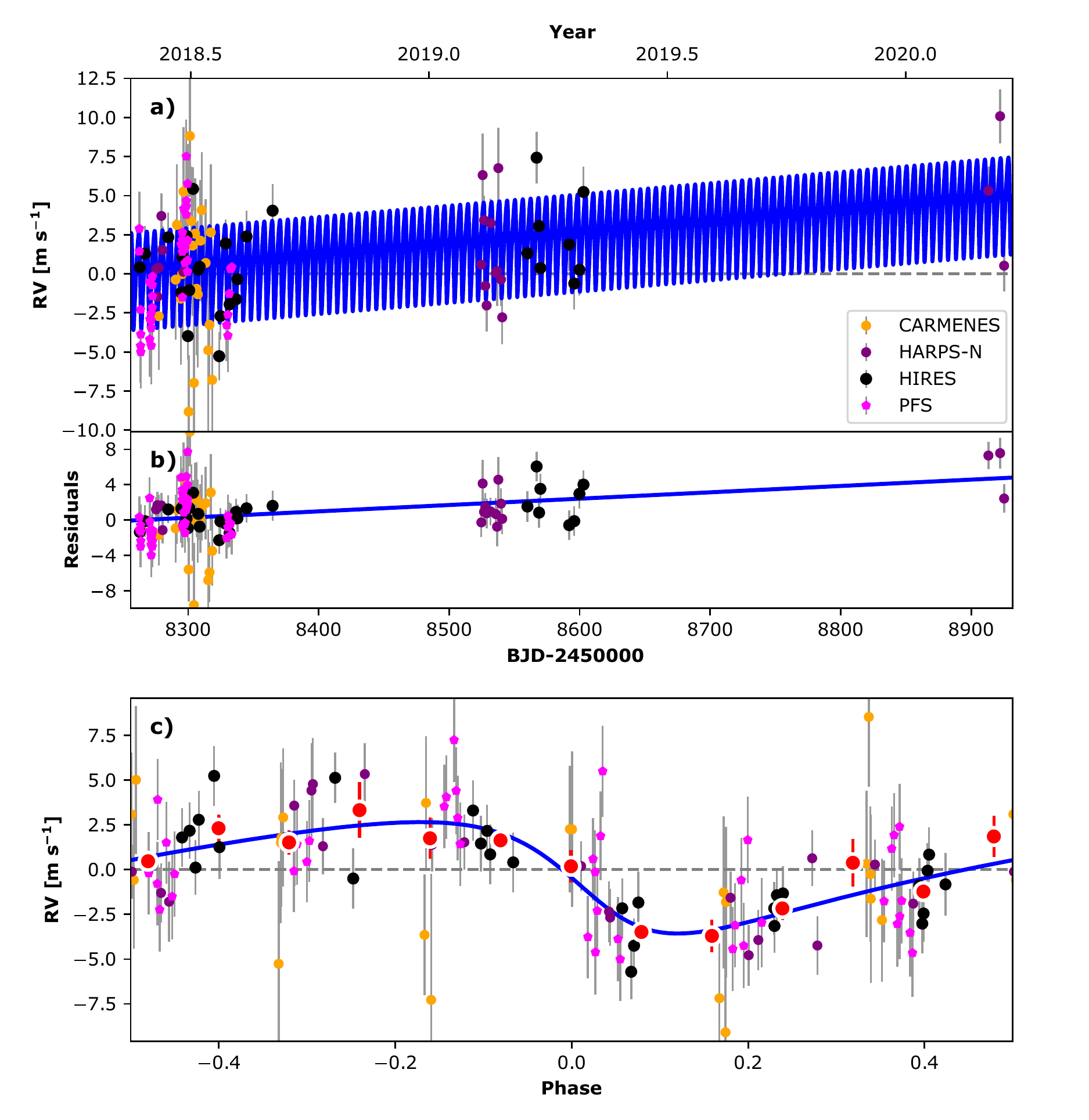}
\vspace{-10pt}
\caption{
 Best-fit 1-planet Keplerian orbital model
  for Wolf 503 (blue line). The maximum likelihood model is plotted while
  the orbital parameters listed in Table \ref{tab:rvparams} are the
  median values of the posterior distributions. We add in quadrature
  the RV jitter terms listed in Table \ref{tab:rvparams} with the
  measurement uncertainties for all RVs.  {\bf b)} Residuals to the
  best fit 1-planet model. {\bf c)} RVs phase-folded
  to the ephemeris of planet b. The small point colors
  and symbols are the same as in panel {\bf a}.  Red circles are the same velocities binned in 0.08 units of orbital
  phase. 
\label{fig:rv_panels}
}
\end{figure*}

We obtained radial velocity measurements of Wolf 503 from four instruments: the Keck Observatory's High Resolution Echelle Spectrometer \citep[HIRES,][]{Vogt1994}, the Calar Alto high-Resolution search for M dwarfs with Exoearths with Near-infrared and optical Echelle Spectrographs \citep[CARMENES,][]{CARMENES,CARMENES18}, the High Accuracy Radial velocity Planet Searcher North \citep[HARPS-N,][]{Cosentino2012}, and the Planet Finder Spectrograph \citep[PFS,][]{Crane2006, Crane2008, Crane2010}. Observations were taken from May 2018 to March 2020 totalling 110 data points. In the following sections we describe the observations and reductions performed for each instrument and the subsequent analysis. The radial velocity points are available in Table \ref{tab:rvs} and displayed in Figure \ref{fig:rv_panels}.

\subsection{PFS Spectroscopy}\label{sec:pfs}
We observed Wolf 503 with PFS from UT 24 May 2018 to UT 03 August 2018 with each exposure totaling 20 minutes producing 42 velocity measurements. The mean internal uncertainty is 1.53~\ms. 

PFS is an iodine cell-based precision RV spectrograph installed on the 6.5m Magellan Clay telescope with an average resolution of $R \simeq$ 130,000. RV values are measured by placing a cell of gaseous I$_2$, which has been scanned with the NIST FTS spectrometer \citep{Nave2017} at a resolution of 1 million, in the converging beam of the telescope. This cell imprints the 5000-6200\AA\ region of the incoming stellar spectra with a dense forest of I$_2$ lines that act as a wavelength calibrator and provide a proxy for the point spread function (PSF) of the spectrometer \citep{MarcyButler1992}. The resulting spectra are split into 2\AA\ chunks, each of which is analyzed using the spectral synthesis technique described in \citet{Butler1996}, which deconvolves the stellar spectrum from the I$_2$ absorption lines and produces an independent measure of the wavelength, instrument PSF, and Doppler shift. The  final Doppler velocity from a given observation is the weighted mean of the velocities of all the individual chunks ($\sim$800 for PFS). The final internal uncertainty of each velocity is the standard deviation of all 800 chunk velocities about that mean.

\subsection{HIRES Spectroscopy}\label{sec:hires}

A total of 27 radial velocity observations of Wolf 503 were obtained from the HIRES spectrograph during the period of 2018 May to 2019 April. HIRES is an iodine (I$_2$) cell-based spectrograph installed on the 10-m Keck I telescope capable of resolutions of $R \simeq$ 50,000 operating between 360-800 nm. Observations were made in collaboration with the California Planet Search (CPS). Spectra were taken with the 14'' by 0.861'' ``C2'' decker with exposures averaging 17 minutes in order to reach the requisite signal to noise ratio of 200 per pixel. We obtained an average signal-to-noise ratio of 223 at 550 nm and an average internal velocity error of 1.08 ${\rm m~s^{-1}}$. Spectra were reduced and radial velocities calculated as described in \cite{Howard2010}.

\subsection{CARMENES Spectroscopy}\label{sec-obs-hrs-carmenes}
We obtained 21 high-resolution spectra of Wolf~503 between June 2018 and July 2018 with the CARMENES instrument mounted on the 3.5\,m telescope at the Calar Alto Observatory, Almer\'ia, Spain, under the observing program S18-3.5-021 (PI Pall\'e). The CARMENES spectrograph has two arms, the visible (VIS) arm covering the spectral range $0.52$--$0.96$\micron~ and a near-infrared (NIR) arm covering the spectral range $0.96$--$1.71$\micron. Here we use only the VIS channel observations to derive radial velocity measurements. All observations were taken with exposure times of 1200\,s resulting in SNR per pixel at 745 nm of CARMENES VIS spectra in the range 41--131. CARMENES performance, data reduction and wavelength calibration are described in \citet{Trifonov18} and \citet{Kaminski18}.

Relative radial velocity values, chromatic index (CRX), differential line width (dLW), and H$\alpha$ index values were obtained using {\tt serval}\footnote{\url{https://github.com/mzechmeister/serval}} \citep{SERVAL}. For each spectrum, we also computed the cross-correlation function (CCF) and its full width half maximum (FWHM), contrast (CTR) and bisector velocity span (BVS) values, following \citet{2020A&A...636A..36L}. The RV measurements were corrected for barycentric motion, secular acceleration and nightly zero-points. Due to the low declination of the star ($\delta=-6.14$ deg), Wolf~503 was observed from Calar Alto at relatively high airmasses (ranging from 1.5 to 2.1), which has a high impact on the telluric contamination of the spectra. Therefore to achieve the highest RV precision, we correct the spectra from telluric absorption using \texttt{Molecfit} \citep{Smette2015,Kausch2015} following the method presented in \citet{2018Sci...362.1388N} and \citet{2018A&A...620A..97S}.

\subsection{HARPS-N Spectroscopy}\label{sec:harpsn}

We collected a total of 20 radial velocity observations of Wolf 503 between June 2018 and March 2020 with the HARPS-N spectrograph installed on the 3.6-m Telescopio Nazionale Galileo (TNG) at the Observatorio del Roque de los Muchachos in La Palma, Spain. These observations were part of the HARPS-N Collaboration's Guaranteed Time Observations (GTO) program. Wolf 503 has an apparent magnitude ${\rm V}$ = 10.28, so we obtained spectra with signal-to-noise ratios in the range SNR = 41 -- 128 (average SNR = 83), at 550 nm in 30 minute exposures, depending on the seeing and sky transparency.  A summary of the observations is provided in Table~\ref{tab:rvs}. The average internal RV error of the observations is 1.19. $\pm$ 0.46 ${\rm m~s^{-1}}$.

The spectra were reduced with version 3.7 of the HARPS-N Data Reduction Software (DRS), which includes corrections for color systematics introduced by variations in seeing \citep{Cosentino2014}. The radial velocities were computed using a numerical weighted mask following the methodology outlined by \citet{Baranne1996}. Masks are chosen based on the closest spectral type of the star and in this case the K5 mask was chosen.

\begin{figure}[h!]
\includegraphics[width=0.46\textwidth]{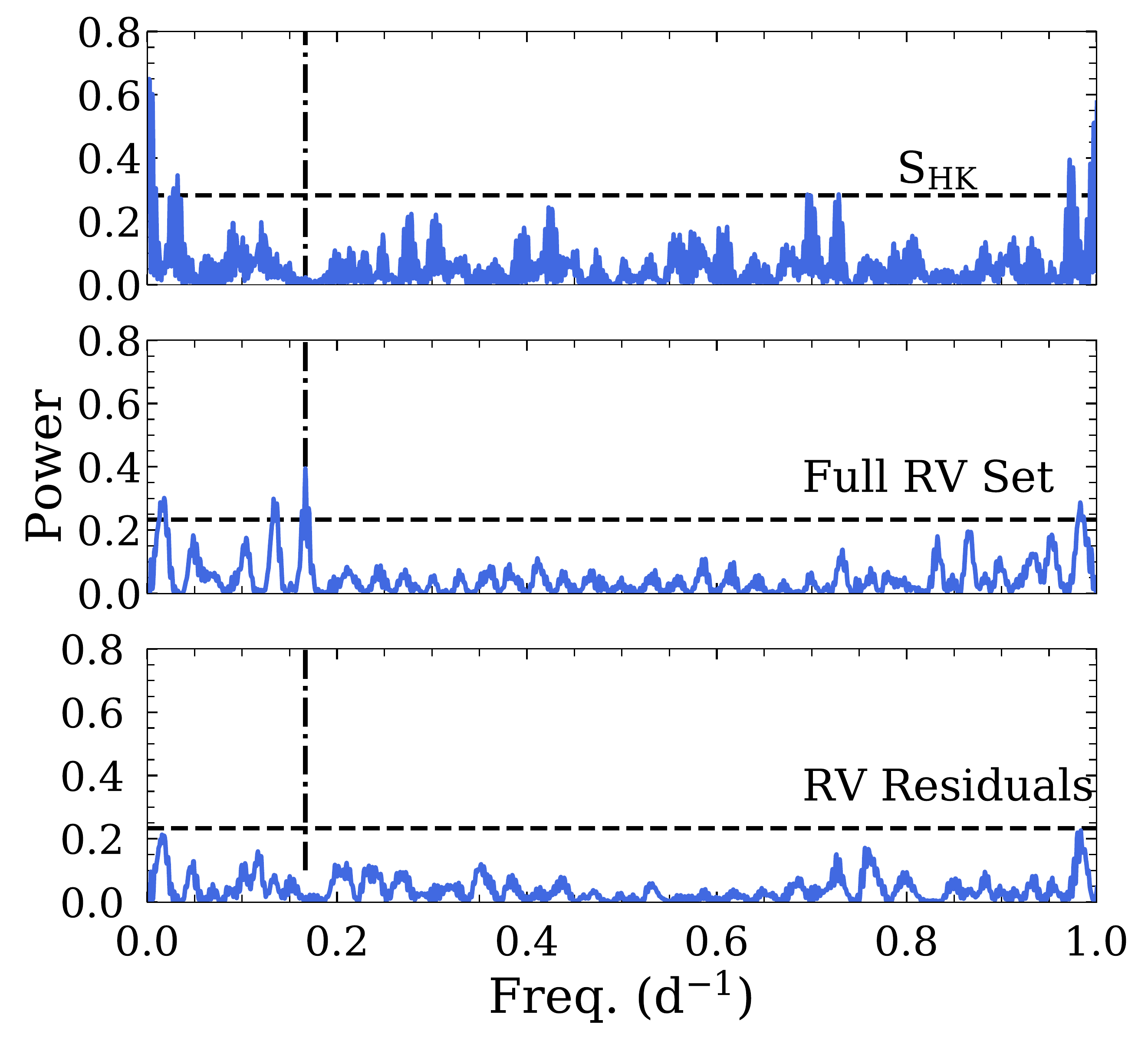}
\caption{Generalized Lomb-Scargle periodograms of all $S_{HK}$ measurements (top), the full radial velocity data set (middle) and the RV residuals after the subtraction of a 1-planet fit (bottom). The dashed horizontal lines represent the power needed to attain a false alarm probability of 0.1\% while the vertical dashed-dotted line marks the period of the planet. Although there are prominent peaks seen in the $S_{HK}$ periodogram these peaks are not seen in the full data set. }
\label{fig:periodogram}
\end{figure}

\subsection{Stellar Activity $\&$ Rotation}\label{sec:activity}

The Mount Wilson $S_{HK}$ index is a commonly used metric of chromospheric activity defined as the ratio of flux in the Ca II H \& K line cores (3968.5\AA~ and 3933.7\AA, respectively) to the flux in the nearby continuum \citep{Wilson1963,Duncan1991}. As part of the California Planet Search (CPS), \cite{Issacson2010} compiled a catalogue of $S_{HK}$ values for over 2,000 stars. A key finding was that K-dwarfs with a color index 1.0 $<$ B-V $<$ 1.3 produce the lowest levels of velocity noise that tend to mimic the radial velocity signals of a planet (known as ``jitter''). Wolf 503 finds itself in this color range with a B-V color of 1.02 suggesting that it is a particularly good radial velocity target. This is confirmed by comparing Wolf 503's $S_{HK}$ index to those of Isaacson and Fischer's sample within that color range. High Resolution Echelle Spectrometer (HIRES) measurements give a median $S_{HK}$ for Wolf 503 of 0.246, much lower than the sample average of 0.536 indicating that Wolf 503 is chromospherically quiet.

\begin{figure}
\centering
\includegraphics[width=0.45\textwidth]{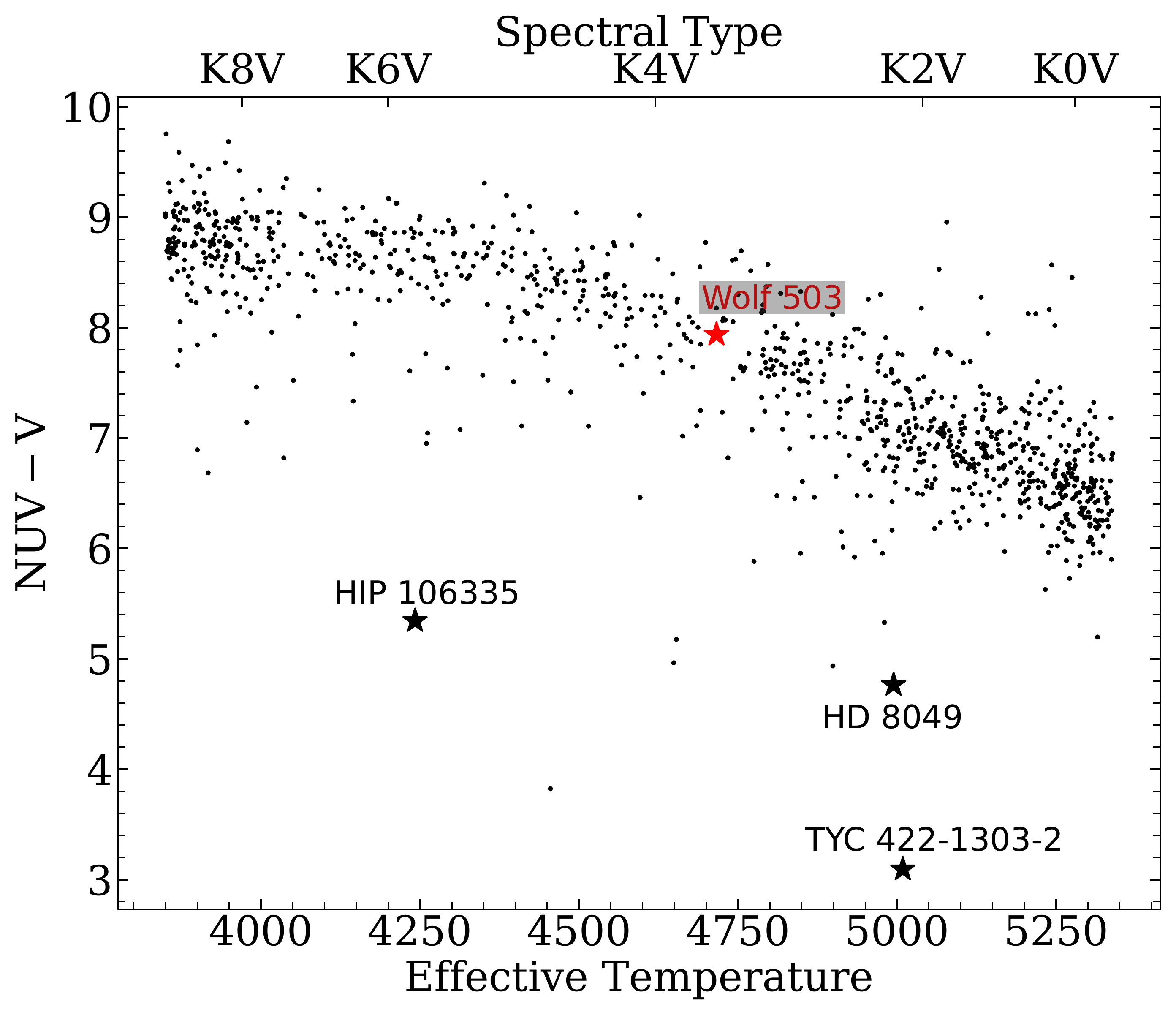}
\vspace{-10pt}
\caption{
   NUV-V colors and effective temperatures for nearby (within 50pc) K-dwarfs from the TESS Input Catalog. Most stars are in a quiescent group signifying low NUV emission while some stars are outliers. A number of these outliers (e.g HIP 106335, HD 8049) are known to have higher stellar activity. Wolf 503 (red star) is a member of the quiescent group which agrees with measurements of stellar activity such as $S_{HK}$.
\label{fig:NUV}
}
\end{figure}

Using Generalized Lomb-Scargle (GLS) periodograms \citep{Zechmeister2009}, we verify that the 6 day signal is present in the radial velocity data (Appendix, Figure \ref{fig:instrument_periodograms}). The planetary signal is prominent in both HIRES and HARPS-N data, but is not clearly seen in the data from PFS or CARMENES.

Due to Wolf 503's low $S_{HK}$, we do not expect stellar activity to impact radial velocity measurements. To verify this, Figure \ref{fig:periodogram} compares Generalized Lomb-Scargle (GLS) periodograms of $S_{HK}$ measurements against the full radial velocity data set (Table \ref{tab:rvs}) to search for stellar activity on the timescale of the orbital period. Data acquisition is described in Section \ref{sec:rv}. The $S_{HK}$ periodogram contains a number of low frequency peaks (below 0.1 d$^{-1}$). The presence of features such as starspots on the stellar surface have the ability to mimic radial velocity signals with periods that can reflect that of the star's rotational period. \cite{Peterson2018} reported a projected rotational velocity of $v\sin{i_*}=$ 0.8 $\pm$ 0.5 \kms. Combining this with the stellar radius of 0.690 $\pm$ 0.02 $R_{\odot}$ we obtain a maximum rotation period of $43 \pm 27$ days. This wide window of possible rotational periods coincides with the low frequency peak structures seen in both the $S_{HK}$ periodogram and RV periodogram (Figure \ref{fig:periodogram}) but these peaks do not coincide.

Another method of assessing the activity of K-dwarf stars is by comparing their high energy flux, in particular near ultra-violet (NUV). At birth, stars have strong magnetic fields and large high energy emissions. As the star ages, a decay in the rotation rate causes a subsequent decrease in this high energy emission. Since this decrease is thought to begin rather quickly, approximately 100 Myrs after formation \citep{RicheyYowell2019}, many K-dwarfs should fall into a quiescent group with low NUV flux. We test this by selecting TESS Input Catalog (TIC) K-dwarfs within 50 pc in the effective temperature range 3850 K $<$ $T_{eff}$ $<$ 5340 K corresponding to spectral types K9V-K0V \citep{Pecaut2013}. This list is then cross-referenced with the stars in the Galaxy Evolution Explorer (\textit{GALEX}) catalog \citep{Bianchi2017} to obtain their NUV magnitudes.

In Figure \ref{fig:NUV} we take the NUV-V colors versus $T_{eff}$ which shows the majority of K-dwarfs have large NUV-V colors forming a quiescent group. NUV-V increases with decreasing stellar temperature, a result that is consistent with a study of M-dwarfs made by \cite{Lepine2013}. We note that Wolf 503, represented as a red star in Fig. \ref{fig:NUV}, has NUV-V = 7.94 placing it within the quiescent group. In contrast, we find a number of K-dwarfs with lower NUV-V colors that could be considered active. Among these few stars, some are known to be active such as HIP 106335, identified to be an ``active/fast rotator'' by \cite{Santos2011}. Additionally, HD 8049 has a high (relative to K-dwarfs, e.g \citealp{Issacson2010}) $S_{HK}$ value of 0.678 \citep{Arriagada2011} and is also found in the active group with an NUV-V = 4.76. Interestingly, some members of the active group, TYC 422-1303-2 among them with the lowest NUV-V, have gone largely unstudied.  

\begin{deluxetable*}{lccc}
\caption{Wolf 503 b Parameters \label{tab:rvparams}}
\tablehead{
  \colhead{Parameter} & 
  \colhead{Name (Units)} & 
  \colhead{Value} &
  \colhead{Provenance}
}
\startdata
\sidehead{\bf{Orbital Parameters}}
$P$ & Period (days) & $6.00127\pm2.1e-5$ & \textit{K2} + \textit{Spitzer} \\
$T\rm{conj}$ & Time of Conjunction (BJD$_{\text{TDB}}$-2450000) & $8191.361449\pm 0.00011$ & \textit{K2} Photometry \\
$a$ & Semi-Major Axis (AU) & $0.05706\pm0.00055$ & \texttt{RadVel} Fit \\
$e$ & Eccentricity & $0.41\pm0.05$  &  Joint RV-Transit Fit\\
$\omega$ & Argument of Periapse (radians) & $1.96\pm0.17$ & \ldots \\
$K$ & Semi-Amplitude (m s$^{-1}$) & $2.98\pm0.36$ & \ldots \\
$b$ & Impact Parameter & $0.65\pm0.06$ & \ldots \\ 
$T_{14}$ & Transit Duration (hours) & $1.33 \pm 0.16$ & \ldots \\
\hline
\sidehead{\bf{Transit Parameters}}
$R_P/R_*$ (\textit{K2}) & Scaled Radius (\%) & $2.791\pm0.049$ & Joint RV-Transit Fit \\
$R_P/R_*$ (\textit{Spitzer}) & Scaled Radius (\%) & $2.73\pm0.13$ & \textit{Spitzer} Photometry \\
$u_0, u_1$ (\textit{K2}) &  Limb Darkening & $\equiv 0.5916,~0.1322$ & \cite{Claret2011} \\
$u_0, u_1$ (\textit{Spitzer}) & Limb Darkening & $\equiv 0.0973,~0.1276$ & \cite{Claret2011} \\
\hline
\sidehead{\bf{Derived Parameters}}
$R$ & Radius (R$_{\oplus}$) & $2.043\pm0.069$ &  \\
$M$ & Mass (M$_{\oplus}$) & $6.26^{+0.69}_{-0.70}$ &  \\
$\rho$ & Density (g cm$^{-3}$) & $2.92^{+0.50}_{-0.44}$ &   \\
$T_{\text{eq}}$ & Equilibrium Temperature (K) & $790\pm15$ & \\
\hline
\sidehead{\bf{Other Parameters}}
$\gamma_{\rm PFS}$ & RV Offset ($\rm m\ s^{-1}$) & $-0.31\pm0.38$ & \texttt{RadVel} Fit  \\
$\gamma_{\rm HIRES}$ & RV Offset ($\rm m\ s^{-1}$) & $-1.23\pm0.38$ & \ldots \\
$\gamma_{\rm HARPS-N}$ & RV Offset ($\rm m\ s^{-1}$) & $−46763.03\pm 0.54$ & \ldots \\
$\gamma_{\rm CARMENES}$ & RV Offset ($\rm m\ s^{-1}$) & $8.14^{+0.94}_{-0.92}$ & \ldots \\
$\dot{\gamma}$ & Acceleration (m s$^{-1}$ d$^{-1}$) & $0.0072^{+0.0016}_{-0.0015} $ & \ldots \\
$\sigma_{\rm PFS}$ & Jitter ($\rm m\ s^{-1}$) & $2.28^{+0.36}_{-0.30}$ &  \ldots\\
$\sigma_{\rm HIRES}$ & Jitter ($\rm m\ s^{-1}$) & $1.34^{+0.37}_{-0.31}$ & \ldots \\
$\sigma_{\rm HARPS-N}$ & Jitter ($\rm m\ s^{-1}$) & $1.42^{+0.45}_{-0.36}$ & \ldots \\
$\sigma_{\rm CARMENES}$ & Jitter ($\rm m\ s^{-1}$) & $3.12^{+0.99}_{-0.82}$ & \ldots \\
$r_1$ & \textit{Spitzer} Ramp Term & $17.3502 \pm 0.0004$ & \textit{Spitzer} Photometry\\
\enddata
\end{deluxetable*}

\subsection{RV-Only Analysis}\label{sec:rv_analysis}

The RV measurements are analyzed using the open source, orbit-fitting tool-kit \texttt{RadVel} \citep[][]{Fulton2018}. With \texttt{RadVel}, a model orbit is fit to the data with the orbital parameters being period ($P$, with a Gaussian prior informed by the value found in \ref{sec:ephemeris}), time of inferior conjunction ($T_{conj}$), RV amplitude ($K$), eccentricity ($e$), and argument of periastron ($\omega$). Other parameters that are fit include an RV offset ($\gamma$) and jitter ($\sigma$) terms for all instruments. During the fitting process, $\sqrt{e}\cos{\omega}$ and $\sqrt{e}\sin{\omega}$ are used in lieu of $e$ and $\omega$ alone in order to avoid biasing the eccentricity.

\begin{deluxetable*}{cccccc}
\centering
\caption{Radial Velocities \label{tab:rvs}}
\tablehead{
  \colhead{Time} & 
  \colhead{RV} & 
  \colhead{RV Unc.} & 
  \colhead{S Index} &
  \colhead{S Index Unc.} &
  \colhead{Inst.} \\
  \colhead{(BJD$_{\text{TDB}}$)} & 
  \colhead{(m s$^{-1}$)} & 
  \colhead{(m s$^{-1}$)} & 
  \colhead{} &
  \colhead{} &
  \colhead{} 
}
\startdata
2458277.49068 & 5.42 & 1.97 & - & - & CARMENES \\
2458290.40436 & 7.76 & 2.62 & - & - & CARMENES \\
2458291.42102 & 11.26 & 2.65 & - & - & CARMENES \\
2458294.43322 & 6.54 & 3.14 & - & - & CARMENES \\
\ldots & \ldots & \ldots & \ldots & \ldots & \ldots \\
2458262.97938 & -0.81 & 1.20 & 0.2385 & 0.001 & HIRES \\
2458266.98198 & 0.07 & 1.36 & 0.2217 & 0.001 & HIRES \\
2458284.78301 & 1.12 & 1.07 & 0.2376 & 0.001 & HIRES \\
2458294.78012 & -2.41 & 0.98 & 0.2372 & 0.001 & HIRES \\
\ldots & \ldots & \ldots & \ldots & \ldots & \ldots \\
2458262.60184 & 2.58 & 0.96 & 0.2929 & 0.0433 & PFS \\
2458262.61853 & 1.12 & 0.87 & 0.2336 & 0.04333 & PFS \\
2458263.53883 & -4.92 & 0.92 & 0.2500 & 0.04344 & PFS \\
2458263.55141 & -2.60 & 0.93 & 0.2370 & 0.04345 & PFS \\
\ldots & \ldots & \ldots & \ldots & \ldots & \ldots \\
2458275.443853 & -46762.739952 & 0.71 & 0.219955 & 0.00189 & HARPS-N \\
2458276.460574 & -46764.503632 & 0.86 & 0.218658 & 0.002924 & HARPS-N \\
2458277.442863 & -46762.6382244 & 0.72 & 0.220267 & 0.001976 & HARPS-N \\
2458279.493334 & -46759.329744 & 0.79 & 0.22194 & 0.002416 & HARPS-N \\
\ldots & \ldots & \ldots & \ldots & \ldots & \ldots \\
\enddata
\tablecomments{Note this table is available in its entirety in a machine-readable form.}
\end{deluxetable*}

Our analysis consists of comparing a simple model of a circular orbit to models with additional parameters such as eccentricity and a linear trend. An MCMC routine is initialized on best-fit values and used to determine the median value of the posterior distribution as well as obtaining an uncertainty for each parameters. As discussed in Section \ref{sec:activity}, neither short-term stellar activity nor rotation are expected to affect our results and so methods of mitigating those effects (e.g. Gaussian processes) are not implemented. We also consider the potential for the Rossiter-Mclaughlin (RM) effect to bias any RV measurements taken during transit. Using Equation 4 from \cite{winn2010}, with the best case scenario of an impact parameter of zero, the maximum amplitude of the RM effect would be $\sim0.6$ ${\rm m~s^{-1}}$; smaller than the average uncertainty for each instrument. In reality, Wolf 503 b  likely has a high impact parameter (see Sec. \ref{sec:rv-trans_analysis}) which renders any bias due to the RM effect even more negligible.

In order to measure the justification of any added parameters we utilized the Akaike information criterion (AIC). An AIC score allows us to compare the goodness of fit of different models while also taking into account over-fitting. The model that minimizes the AIC is considered optimum. The difference between the lowest AIC and the AIC of a model in question ($\Delta$AIC = $\text{AIC}_{model}-\text{AIC}_{min}$) allows us to reject models that either poorly describe the data or contain too many parameters:  $\Delta$AIC $<$ 2 shows little difference between the two models, 2 $<$ $\Delta$AIC $<$ 10 indicates less support for the model, and a $\Delta$AIC $>$ 10 means the model is strongly disfavored.

When comparing models of circular and eccentric orbits both with and without acceleration terms we find that an eccentric orbit with a linear trend is by far the preferred model with a circular orbit being disfavored ($\Delta$AIC = 9.57) and any model without a trend included being entirely ruled out ($\Delta$AIC = 17). Our analysis of the radial velocity data alone reveals an orbital eccentricity of $0.35 \pm 0.09$. The discrepancy between this value and the one found from the \textit{K2} photometry is addressed with a joint RV-transit fit in Section \ref{sec:rv-trans_analysis}. 

A trend in RV data can indicate the presence of a long-period, massive companion however they can also be caused by long term stellar activity. We observe positive trends in both the S-index and Full Width at Half Maximum values from HIRES and HARPS-N suggesting that this trend is stellar in origin rather than evidence of planet `c'. However, we also note that there exists only a slight correlation between the S-index values and the radial velocity measurements with a Pearson correlation coefficient of 0.26. Further monitoring of this system is likely needed to determine the nature of this trend.

\subsection{Joint RV-Transit Analysis} \label{sec:rv-trans_analysis}

The discrepancy between the eccentricity values predicted from the photo-eccentric modeling of the \textit{K2} photometry and from the RV data alone suggests that a joint RV-transit analysis may be necessary for Wolf 503 b. Often, the degeneracy between the impact parameter and eccentricity can result in small estimates of $b$ \citep[][]{Dawson2012}. We attempt to resolve this discrepancy by modeling the photometry and radial velocity measurements simultaneously. 

Our joint model is constructed using \texttt{exoplanet} using the same parameters from the photo-eccentric and radial velocity models. Priors were placed on $\rho_*$ using the values in Table \ref{tab:stellarparams}. Without the orbital information we gain from the RV analysis, the impact parameter derived from photometry is both small and unconstrained at $b=0.18\pm0.11$ but our joint model revises this value to $b=0.65\pm0.06$ and produces a new, slightly higher eccentricity estimate of $e=0.41\pm0.05$. The scaled planet radius is also affected, due to the dependence on both $b$ and $e$, increasing to $2.79\pm0.05\%$. All other parameters remained consistent with the values found with either the photo-eccentric model or RV only model. A summary of model and derived parameters is provided in Table \ref{tab:rvparams}.

\section{\textbf{Discussion}} \label{sec:discussion}

From our radial velocity analysis, we find Wolf 503 b has a mass of \mass \umass \mearth  and, combining this with a radius of  \radius \uradius~R$_{\Earth}$, has a bulk density of \density \udensity \densityunits. These measurements allow us to place this planet in context and investigate its viability as a target for atmospheric characterization.

\subsection{Interior Models and Formation Theories}\label{interiors}

Sub-Neptunes are typically described as low-density planets with modest H-He envelopes making up anywhere between 0.1-10\% of the planet's mass. Super-Earths, on the other hand, are thought to be smaller planets with higher densities stripped bare of any envelope. With the newly acquired mass of Wolf 503 b we employ the Structure Model Interpolator tool \citep[\texttt{smint},][]{Piaulet2020} to determine the envelope mass fractions for both a \water dominated planet and one with an H-He envelope. \texttt{smint} uses the model grids of \citealp{Lopez2014} and \citealp{Zeng2016} to determine the mass fraction for H-He and water, respectively. The former model grids assume a core composed of a 2:1 mix of rock and Iron while the latter employs a 2-layer model reflective of Earth's core and mantle. We find that Wolf 503 b is entirely consistent with an Earth-like core of rock and iron with either an \water\ mass fraction of \waterfracshort \uwaterfracshort\% or an H/He mass fraction ($f_{H,He}$) of \hhefrac \uhhefrac\%. These values are consistent with common definitions of sub-Neptunes.

The lack of sub-Neptunes orbiting the Sun means that we still have much to discover about their origins. Early investigations into planet formation focused on replicating the system architecture of our own Solar System and, even though the planets in our system can be formed at their current positions, it is generally accepted that this is not feasible for hot/warm sub-Neptunes through classic core-accretion \citep{Bodenheimer2014,Inamdar2015, Venturini2020}. \cite{Schlichting2014} calculated the required enhancement of the minimum mass solar nebula (MMSN) needed for in-situ formation of sub-Neptunes for various masses and at varying distances from its host star. For a roughly $5~M_{\Earth}$ planet forming at 0.05 AU, the MMSN would need to contain 90 times more solid material in the inner-disk. If the metallicity of the host star is reflective of the refractory content of the proto-planetary disk then this enhancement requirement is especially unreasonable for Wolf 503 whose metal content is only 30\% that of the Sun. Pebble accretion could offer an \textit{in-situ} formation pathway for Wolf 503 b, however, pebble accretion may tend to form systematically drier planets since the pebbles should lose most of their volatiles during their journey to the inner disk \citep{Oka2011, Ida2019}. Planetesimals that form beyond the snow line are more likely to retain their volatiles and can contain 10--50\% water by mass \citep{Izidoro2019} resulting in vastly different compositions for migrating and \textit{in-situ} planets. Although planetary compositions derived from bulk density alone are degenerate, the high bulk water composition of Wolf 503 b could imply a formation beyond the snowline and subsequent migration inwards.

\begin{figure*}[t!]
\centering
\includegraphics[height=8.0in,width=7.0in,keepaspectratio]{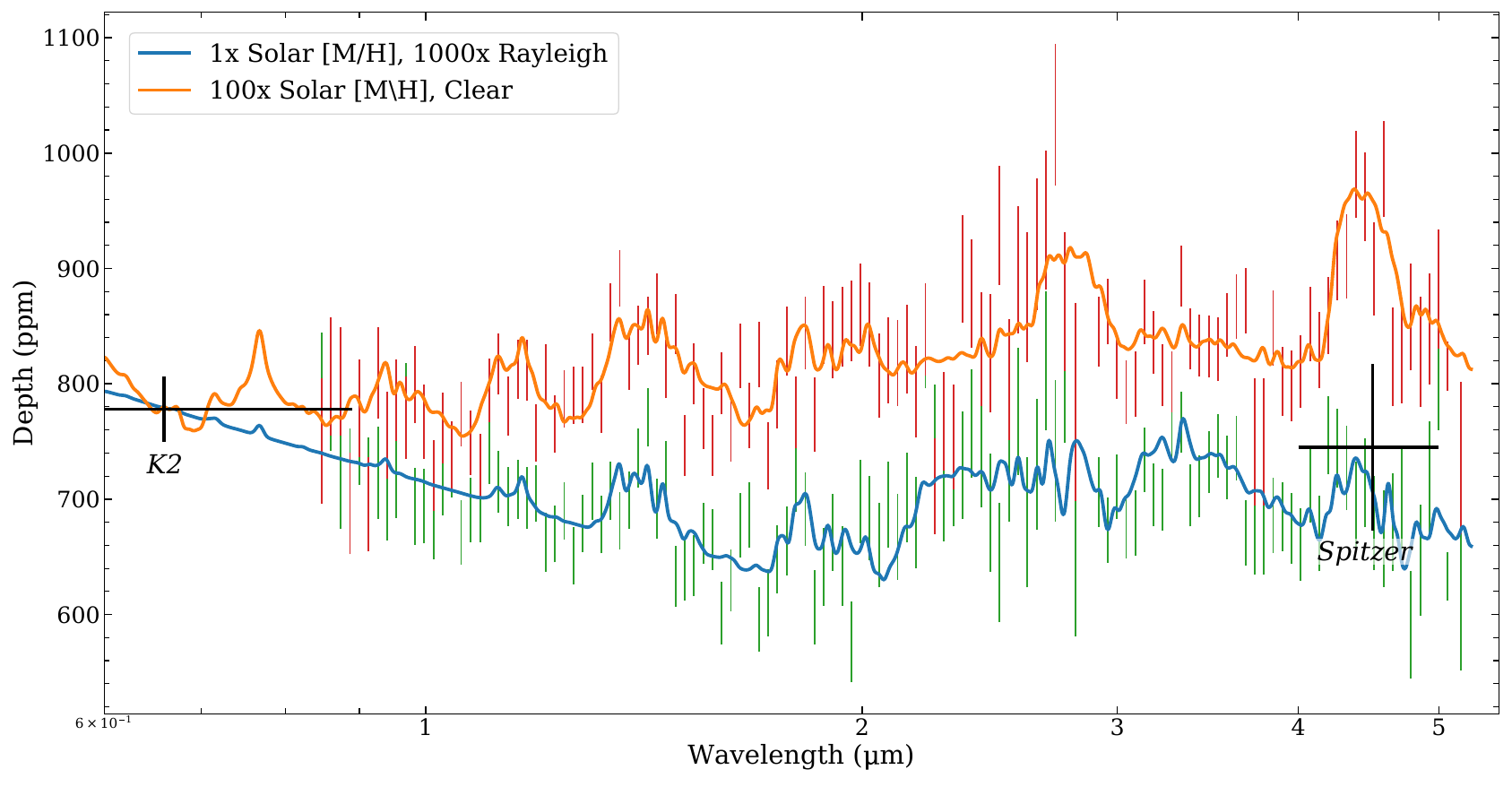}
\vspace{-10pt}
\caption{
   A selection of two atmospheric model spectra of the eight that were compared. The lower spectrum shows a 1x solar [M/H] with strong haze effects while the one above is an atmosphere with higher mean molecular weight but without the effect of aerosols. Spectral features are clearly discernible in both.
\label{fig:wolf503b_atmo}
}
\end{figure*}

When considering planets of high $T_{eq}$, it is also important to note that the planet we characterize today has evolved significantly since its formation. Planets hosted by relatively long-lived stars can provide insight into the end products of mass loss mechanisms such as photoevaporation \citep{OwenWu2013} and core-powered mass loss. Wolf 503 b, orbiting an $11 \pm 2$ Gyr old K-dwarf, likely experienced appreciable photoevaporation of its atmosphere. Neptune-class planets (M $\sim 20M_{\Earth}$) can have the majority of their atmosphere removed by its host star; in the most extreme cases the planet is left with H/He envelopes of fractions of a percent consistent with the H/He mass fractions found for Wolf 503 b. Much of the evaporation is thought to occur in the first 100 Myrs during a ``saturation'' phase early in the star's life when X-ray emission is at its peak and independent of rotation period \citep{Owen2017}. Given the age of this system, Wolf 503 b could be an example of the end product of photoevaporation.

\subsection{Potential for Atmospheric Characterization}\label{atmo}

Equipped with a precise mass measurement of Wolf 503 b, we are now able to more carefully consider the viability of this planet for transmission spectroscopy. A calculation of the transmission spectroscopy metric (TSM) of \cite{Kempton2018} places Wolf 503 b (TSM=63.9) in the top twenty best atmospheric follow-up targets in the size range $1R_{\Earth}<R_{P}<4R_{\Earth}$ \citep{Guo2020}. This immediately suggests Wolf 503 b as a potential target for atmospheric characterization. 

A mass uncertainty below 20\% decouples the similar effects that both high surface gravity and a high mean molecular weight composition have on atmospheric spectra allowing us to investigate the latter. However, with that degeneracy broken, we are potentially faced with another. The tentative correlation of the water absorption amplitude with $T_{eq}$ \citep{Crossfield2017} suggest that hazes should be important considerations when modeling the atmospheres of planets like Wolf 503 b. A $T_{eq}$ of 790K places Wolf 503 b in a region where hazes might be commonplace for warm Neptunes, but its small size and low H/He mass fraction could indicate enhanced metallicity \cite{Fortney2013, Venturini2016}. Both of these factors can have similar, flattening effects on transmission spectra.

We consider the ability of the James Webb Space Telescope (JWST) to distinguish between these effects by generating model spectra using \texttt{ExoTransmit} \citep{Kempton2017} with varying degrees of aerosols ranging from a clear atmosphere to hazes with 100x and 1000x solar Rayleigh scattering or a cloud deck at 0.01 bar. For each of these aerosol compositions we simulated metallicities of 1x and 100x solar [M/H]. These spectra were then used to simulate JWST observations using  NIRISS (Single object slitless spectroscopy covering $0.6-2.8 \mu m$) and NIRSpec (Bright object time series with G395H covering $2.87-5.27\mu m$) instrument modes. Simulations were made with \texttt{PandExo} \citep{Batalha2017b} assuming a resolution of $R=35$, a baseline equal in time to that of the transit and a zero noise floor. The model spectra were then smoothed to match the native resolution of the instrument and binned down to match the resolution of the simulated observations. Using a weighted least squares routine, each simulated JWST spectrum was then fit with both a linear model and the model spectra (including the model the simulation was generated from). The corresponding reduced $\chi^{2}$ statistics and p-values (summarized in Tables \ref{tab:niriss} \& \ref{tab:nirspec}) were calculated and used to compare the models. 

The first question one would ask is whether these atmospheres are detectable (i.e is the linear model strongly rejected?). In this analysis we will consider a p-value $>$ 0.05 to be a non-rejection of the model being fitted (or a non-detection in the linear case), 0.05 $>$ p $>$ 0.006 to be a weak rejection, and a p $<$ 0.006 to be a strong rejection of the model. For NIRISS, each set of simulated spectra show an unambiguous detection with the exception of the cloud deck spectra for both metallicities (Figure \ref{fig:wolf503b_atmo}). This is not too surprising; the presence of clouds is expected to be a significant challenge when studying exoplanet atmospheres. Although, it is interesting to note that the cloudy 100x solar spectrum was a weak detection whereas its 1x solar counterpart was indistinguishable from the linear model. One would expect the combined effect of a cloudy, high mean molecular weight atmosphere would result in a stronger rejection than one of lower metallicity. For NIRSpec, the situation is slightly less optimistic. An aerosol-free composition was the only low metallicity atmosphere detectable but on the other hand, \textit{all} high metallicity atmospheres were detectable. 

Among the models that were detectable, we then ask whether these models are differentiable from one another. Both NIRISS and NIRSpec will be capable of distinguishing between atmospheres of different metallicity but NIRISS will be particularly useful for detecting possible hazes which is consistent with the results found by \cite{Batalha2017}. For a solar metal content, NIRISS was able to resolve the differences in spectra due to various strengths of Rayleigh scattering, however for a metallicity 100 times solar the ability to detect these differences was lost.  Cloud decks at lower pressures (higher altitude) would likely exacerbate this issue and, although not investigated here, we also have no reason to assume exoplanet atmospheres cannot contain both hazes and clouds potentially muting the effect of Rayleigh scattering. 

From the TSM alone, Wolf 503 b proves to be a strong candidate for further atmospheric characterization. Our analysis shows that, at the very least, we could expect to differentiate a low mean molecular weight atmosphere from a higher one. Evidence of aerosols is also well within reach of JWST with a distinction between hazes and clouds being possible if the atmosphere has a close to solar metal content. Considering the increase in information to be gained from a low metallicity atmosphere, the relative metal-poorness of Wolf 503 b's host star only solidifies further this planet's potential as a follow-up target. Forming from a metal-poor disk may be helpful to keep the subsequent metallicity of the atmosphere low as well.

\section{Conclusions}

In this paper we characterized the sub-Neptune Wolf 503 b. Through radial velocity measurements we find that it is on an eccentric orbit ($e=0.41\pm0.05$) and determine its mass to be \mass \umass \mearth. Employing stellar activity indicators, we find that the host star is indeed a well-behaved K-dwarf furthering this spectral class' reputation as the most amenable to radial velocity studies. We also compare Wolf 503 b to other K-dwarfs with recorded NUV measurements from the \textit{GALEX} survey and find that it is a member of a large group with low NUV emission. 

A joint analysis of previously unused short cadence \textit{K2} photometry and radial velocity data in combination with \textit{Gaia} EDR3 data, provided us with a radius of \radius \uradius \rearth resulting in a bulk density of \density \udensity$\text{ gcm}^{-3}$. This low density helps confirm Wolf 503 b as a sub-Neptune with either a substantial \water mass fraction of \waterfracshort \uwaterfracshort \% or an H-He mass fraction of \hhefrac \uhhefrac\%.

To enable future investigations of this planet, we utilized a \textit{Spitzer} transit to further constrain ephemerides providing accurate transit predictions well into the JWST era. This analysis resulted in a 5-fold reduction in transit time uncertainty as compared to predictions made with values from \citet{Peterson2018}.

We also explore the possibility of detecting a high metallicity atmosphere in addition to hazes finding that, in agreement with previous work by \citet{Batalha2017}, that the NIRISS instrument will be an indispensable tool for atmospheric studies of Sub-Neptunes. The presence of clouds or the combination of strong haze effects with a high metallicity atmosphere understandably makes measurements less conclusive. We have found that Wolf 503 b offers itself as a good candidate for JWST follow-up observations and can act as a case-study for planets orbiting old, metal-poor stars.

\acknowledgments

The authors thank the anonymous referee whose thorough review greatly increased the quality of this publication. We also thank the time assignment committees of the University of California, the California Institute of Technology, NASA, and the University of Hawaii for supporting the TESS-Keck Survey with observing time at Keck Observatory and on the Automated Planet Finder.  We thank NASA for funding associated with our Key Strategic Mission Support project.  We gratefully acknowledge the efforts and dedication of the Keck Observatory staff for support of HIRES and remote observing. 
We recognize and acknowledge the cultural role and reverence that the summit of Maunakea has within the indigenous Hawaiian community. We are deeply grateful to have the opportunity to conduct observations from this mountain.

Based on observations made with the Italian Telescopio Nazionale Galileo (TNG) operated on the island of La Palma by the Fundacion Galileo Galilei of the INAF (Istituto Nazionale di Astrofisica) at the Spanish Observatorio del Roque de los Muchachos of the Instituto de Astrofisica de Canarias.

This paper includes data gathered with the 6.5 meter Magellan Telescopes located at Las Campanas Observatory, Chile.

The HARPS-N project has been funded by the Prodex Program of the Swiss Space Office (SSO), the Harvard University Origins of Life Initiative (HUOLI), the Scottish Universities Physics Alliance (SUPA), the University of Geneva, the Smithsonian Astrophysical Observatory (SAO), and the Italian National Astrophysical Institute (INAF), the University of St Andrews, Queen's University Belfast, and the University of Edinburgh.

Part of this research was carried out at the Jet Propulsion Laboratory, California Institute of Technology, under a contract with the NationalAeronautics and Space Administration (NASA)

This work is based in part on observations made with
the Spitzer Space Telescope, which was operated by the
Jet Propulsion Laboratory, California Institute of Technology under a contract with NASA. Support for this
work was provided by NASA through an award issued
by JPL/Caltech.

This work has made use of data from the European Space Agency (ESA) mission
{\it Gaia} (\url{https://www.cosmos.esa.int/gaia}), processed by the {\it Gaia}
Data Processing and Analysis Consortium (DPAC,
\url{https://www.cosmos.esa.int/web/gaia/dpac/consortium}). Funding for the DPAC
has been provided by national institutions, in particular the institutions
participating in the {\it Gaia} Multilateral Agreement.

Annelies Mortier acknowledges support from the senior Kavli Institute Fellowships.

This project has received funding from the European Research Council (ERC) under the European Union’s Horizon 2020 research and innovation programme (grant agreement SCORE No 851555)

This research made use of Lightkurve, a Python package for Kepler and TESS data analysis \citep{Lightkurve2018}.

\facilities{Keck:I(HIRES), Magellan:Clay(PFS), Observatorio  del Roque de los Muchachos:TNG(HARPS-N),  Calar Alto:Zeiss 3.5m(CARMENES), Spitzer} 

\software{exoplanet \citep{Foreman-Mackey2021}, radvel \citep{Fulton2018}, Lightkurve \citep{Lightkurve2018}, batman \citep{Kreidberg2015}, isoclassify \citep{Huber2017}, emcee \citep{Foreman2013}, PandExo \citep{Batalha2017b}, ExoTransmit \citep{Kempton2017}, smint \citep{Piaulet2020}}

\newpage

\appendix

\section{Per-Instrument Periodograms}\label{sec:instrument_periodograms}
\begin{figure}[h!]
\centering
\includegraphics[width=0.5\textwidth]{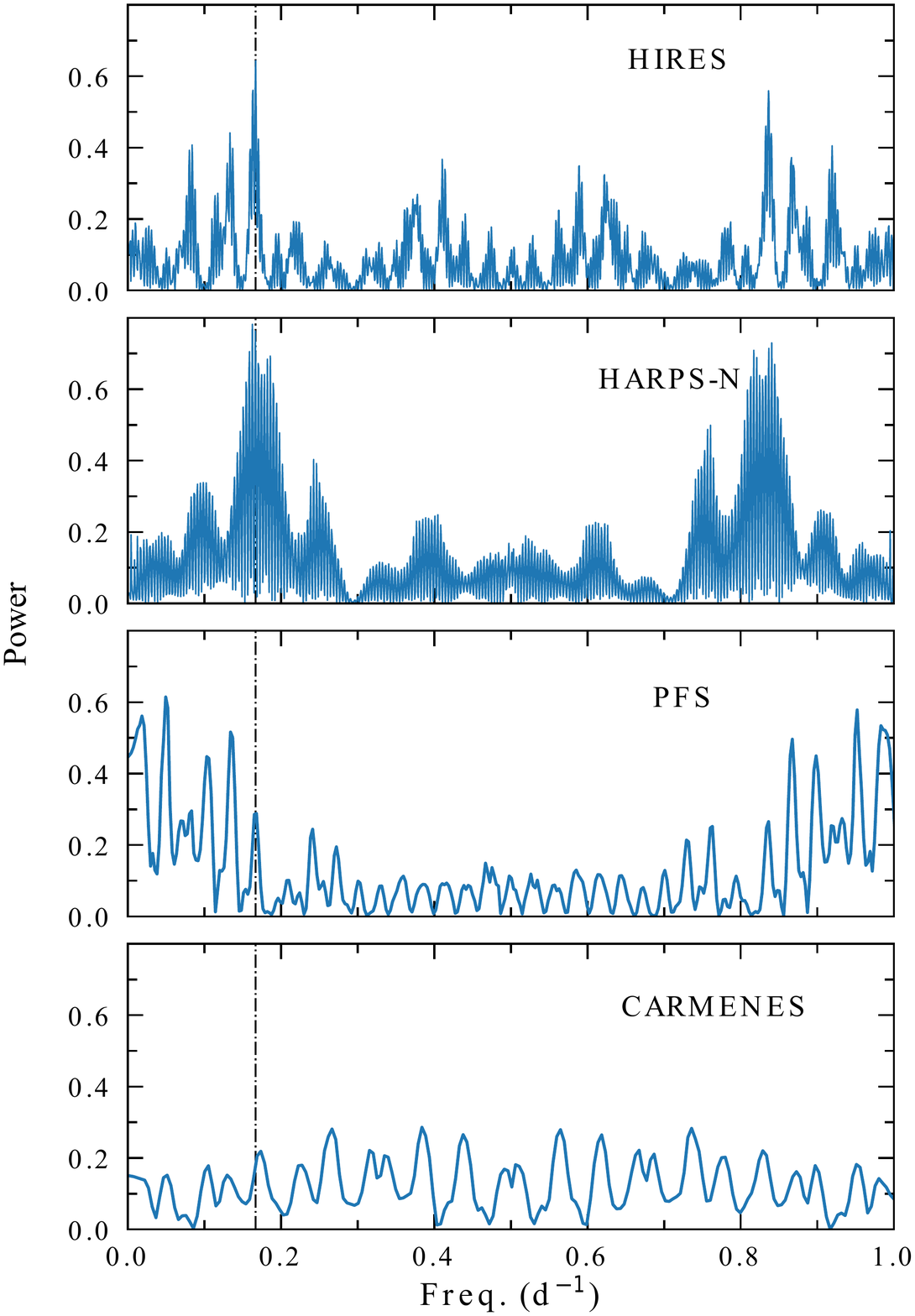}
\caption{Generalized Lomb-Scargle periodograms of all four instruments used to measure Wolf 503's radial velocity signal. The orbital frequency due to Wolf 503 b is prominently seen in the periodogram for both HIRES and HARPS-N but not in the data sets for both PFS and CARMENES. The dash-dotted line represents the orbital frequency of Wolf 503 b ($\sim$0.167 d$^{-1}$)}
\label{fig:instrument_periodograms}
\end{figure}

\section{Atmospheric Model Comparison}

\begin{sidewaystable*}[ph!]
\setlength{\tabcolsep}{0.0cm}
 \centering
 \caption{NIRISS Fitting Results. Values displayed are the resultant $\chi^2$'s with each cell color-coded according to the p-value. \\
 Blue: p$>$0.05, Light Red: 0.05$>$p$>$0.006, Dark Red: p$<$0.006. \label{tab:niriss}}
 \renewcommand{\arraystretch}{1.5}
 \begin{tabular}{cc|P{1.5cm}P{2.3cm}P{2.3cm}P{2.3cm}||P{1.5cm}P{2.3cm}P{2.3cm}P{2.3cm}P{1.7cm}}

 \hline
 \multicolumn{2}{c|}{\multirow{2}{*}{\diagbox[outerrightsep=-3pt,innerleftsep=-1.1cm,innerrightsep=-0.1cm,width=3.2cm]{Fitting Model}{JWST Model}}} & \multicolumn{4}{c||}{1x [M/H]} & \multicolumn{4}{c}{100x [M/H]} & \multicolumn{1}{c}{} \\
 & & Clear & 100x Rayleigh & 1000x Rayleigh& 0.01 Bar & Clear & 100x Rayleigh& 1000x Rayleigh& 0.01 Bar \\
 \hline
 \parbox[t]{2mm}{\multirow{4}{*}{\rotatebox[origin=c]{90}{1x [M/H]}}} 
 
 & Clear & \cellcolor{blue!25}87.8 & 165.62 \cellcolor{red!50}& \cellcolor{red!50}356.3 & \cellcolor{red!50} 204.2& \cellcolor{red!50}165.2& \cellcolor{red!50}163.9 & \cellcolor{red!50}164.1 & \cellcolor{red!50}185.4  \\
 
 & 100x Rayleigh & \cellcolor{red!50}160.0 & \cellcolor{red!50}112.4 & \cellcolor{red!50}159.0 & \cellcolor{red!50}132.1 & \cellcolor{red!50}113.5 & \cellcolor{red!50}112.1& \cellcolor{red!50}111.2& \cellcolor{red!50}114.5  \\
 
 & 1000x Rayleigh & \cellcolor{red!50}280.9 & \cellcolor{red!50}138.3 & \cellcolor{blue!25}85.2 & \cellcolor{red!50}138.36 & \cellcolor{red!50}177.6 & \cellcolor{red!50}176.7 & \cellcolor{red!50}174.1 & \cellcolor{red!50}162.9  \\
 
 & 0.01 Bar & \cellcolor{red!50}212.0 & \cellcolor{red!50}137.1 & \cellcolor{red!50}131.2 & \cellcolor{blue!25}80.2 & \cellcolor{blue!25}89.6 & \cellcolor{blue!25}87.3 & \cellcolor{blue!25}82.9& \cellcolor{blue!25} 75.2  \\
 \hline
 \hline
 \parbox[t]{2mm}{\multirow{4}{*}{\rotatebox[origin=c]{90}{100x [M/H]}}} 
 
 & Clear & \cellcolor{red!50}172.8 & \cellcolor{red!50}153.07 & \cellcolor{red!50}208.6 & \cellcolor{red!25}107.8 & \cellcolor{blue!25}90.1 & \cellcolor{blue!25}89.7 & \cellcolor{blue!25}88.6 & \cellcolor{blue!25}86.8  \\

 & 100x Rayleigh & \cellcolor{red!50}214.1 & \cellcolor{red!50}168.5 & \cellcolor{red!50}200.5 & \cellcolor{red!50}126.6 & \cellcolor{red!50}117.2 & \cellcolor{red!50}117.1 & \cellcolor{red!50}116.2 & \cellcolor{red!50}113.6  \\
 
 & 1000x Rayleigh & \cellcolor{red!50}160.2 & \cellcolor{red!50}115.4 & \cellcolor{red!50}159.5 & \cellcolor{red!25}100.9 & \cellcolor{blue!25}75.1 & \cellcolor{blue!25}75.0 & \cellcolor{blue!25}74.2 & \cellcolor{blue!25}74.0  \\
 
 & 0.01 Bar & \cellcolor{red!50}178.5 & \cellcolor{red!50}113.2 & \cellcolor{red!50}128.9 & \cellcolor{blue!25}91.2 & \cellcolor{blue!25}82.4 & \cellcolor{blue!25}78.9 & \cellcolor{blue!25}77.7 & \cellcolor{blue!25}75.7  \\
 \hline
 \hline
 & Linear &  \cellcolor{red!50}237.3 & \cellcolor{red!50}171.7 & \cellcolor{red!50}137.4 & \cellcolor{blue!25} 80.4 & \cellcolor{red!50}126.6 & \cellcolor{red!50}130.2 & \cellcolor{red!50}131.6 & \cellcolor{red!25}104.9 \\
 \hline
 \end{tabular}
\end{sidewaystable*}

\begin{sidewaystable*}[ph!]
\setlength{\tabcolsep}{0.0cm}
 \centering
 \caption{NIRSpec Fitting Results. Values displayed are the resultant $\chi^2$'s with each cell color-coded according to the p-value. \\
 Blue: p$>$0.05, Light Red: 0.05$>$p$>$0.006, Dark Red: p$<$0.006.\label{tab:nirspec}}
 \renewcommand{\arraystretch}{1.5}
 \begin{tabular}{cc|P{1.5cm}P{2.3cm}P{2.3cm}P{2.3cm}||P{1.5cm}P{2.3cm}P{2.3cm}P{2.3cm}P{1.7cm}}

 \hline
 \multicolumn{2}{c|}{\multirow{2}{*}{\diagbox[outerrightsep=-3pt,innerleftsep=-1.1cm,innerrightsep=-0.1cm,width=3.2cm]{Fitting Model}{JWST Model}}} & \multicolumn{4}{c||}{1x [M/H]} & \multicolumn{4}{c}{100x [M/H]} & \multicolumn{1}{c}{} \\
 & & Clear & 100x Rayleigh & 1000x Rayleigh& 0.01 Bar & Clear & 100x Rayleigh& 1000x Rayleigh& 0.01 Bar \\
 \hline
 \parbox[t]{2mm}{\multirow{4}{*}{\rotatebox[origin=c]{90}{1x [M/H]}}} 
 
 & Clear & \cellcolor{blue!25}46.7 & 41.7 \cellcolor{blue!25}& \cellcolor{blue!25}40.3 & \cellcolor{blue!25} 42.0& \cellcolor{red!50}130.7 & \cellcolor{red!50}128.4 & \cellcolor{red!50}127.3 & \cellcolor{red!50}118.1 \\
 
 & 100x Rayleigh & \cellcolor{blue!25}34.5 & \cellcolor{blue!25}34.2 & \cellcolor{blue!25}34.3 & \cellcolor{blue!25}33.1 & \cellcolor{red!50}133.7 & \cellcolor{red!50}124.6 & \cellcolor{red!50}120.5 & \cellcolor{red!50}112.9 \\
 
 & 1000x Rayleigh & \cellcolor{blue!25}38.2 & \cellcolor{blue!25}36.3 & \cellcolor{blue!25}28.7 & \cellcolor{blue!25}26.0 & \cellcolor{red!50}88.3 & \cellcolor{red!50}80.8 & \cellcolor{red!50}80.8 & \cellcolor{red!50} 73.6 \\
  
 & 0.01 Bar & \cellcolor{blue!25}29.0 & \cellcolor{blue!25}28.4 & \cellcolor{blue!25}28.1 & \cellcolor{blue!25}27.5 & \cellcolor{red!50}119.1 & \cellcolor{red!50}117.6 & \cellcolor{red!50}115.1 & \cellcolor{red!50}101.3 \\
 
 \hline
 \hline
 \parbox[t]{2mm}{\multirow{4}{*}{\rotatebox[origin=c]{90}{100x [M/H]}}} 
 
 & Clear & \cellcolor{red!50}93.5 & \cellcolor{red!50}93.1 & \cellcolor{red!50}90.9 & \cellcolor{red!50}81.6 & \cellcolor{blue!25}35.0 & \cellcolor{blue!25}35.0 & \cellcolor{blue!25}34.8 & \cellcolor{blue!25}32.0 \\

 & 100x Rayleigh & \cellcolor{red!50}103.7 & \cellcolor{red!50}103.2 & \cellcolor{red!50}99.5 & \cellcolor{red!50}89.4 & \cellcolor{blue!25}27.5 & \cellcolor{blue!25}27.5 & \cellcolor{blue!25}26.9 & \cellcolor{blue!25}26.1 \\
 
 & 1000x Rayleigh & \cellcolor{red!50}128.9 & \cellcolor{red!50}128.9 & \cellcolor{red!50}126.1 & \cellcolor{red!50}111.8 & \cellcolor{blue!25}31.4 & \cellcolor{blue!25}31.3 & \cellcolor{blue!25}30.1 & \cellcolor{blue!25}30.0 \\
 
 & 0.01 Bar & \cellcolor{red!50}157.8 & \cellcolor{red!50}157.7 & \cellcolor{red!50}154.2 & \cellcolor{red!50}137.7 & \cellcolor{blue!25}38.4 & \cellcolor{blue!25}38.1 & \cellcolor{blue!25}38.0 & \cellcolor{blue!25}36.4  \\
 \hline
 \hline
 & Linear & \cellcolor{red!50}74.5 & \cellcolor{blue!25}38.4 & \cellcolor{blue!25} 31.5 & \cellcolor{blue!25}38.4 & \cellcolor{red!50}53.9 & \cellcolor{red!50}71.9 & \cellcolor{red!50}69.4 & \cellcolor{red!50}86.1 \\
 
\hline
\end{tabular}
\end{sidewaystable*}

\newpage

\bibliography{main}
\bibliographystyle{aasjournal}

\end{document}